\documentclass[fleqn,usenatbib]{mnras}
\usepackage{CJKutf8}
\newcommand{\CJKname}{\begin{CJK*}{UTF8}{bkai}黃鎔鈞\end{CJK*}} 

\usepackage{newtxtext,newtxmath}

\usepackage[T1]{fontenc}

\DeclareRobustCommand{\VAN}[3]{#2}
\let\VANthebibliography\thebibliography
\def\thebibliography{\DeclareRobustCommand{\VAN}[3]{##3}\VANthebibliography}

\usepackage{graphicx}	
\usepackage{amsmath}	
\usepackage{bm} 

\usepackage{multirow}
\usepackage{diagbox}
\usepackage
{caption}
\usepackage{subcaption}

\usepackage{orcidlink}
\usepackage{xcolor}

\usepackage{hyperref}
\usepackage{appendix}

\usepackage{tabularx}  

\newcommand{\citeg}[1]{\citep[e.g.,][]{#1}}

\defcitealias{calzetti00}{Calzetti}
\defcitealias{chabrier03}{Chabrier}
\defcitealias{baldwin81}{BPT}
\defcitealias{dopita16}{D16}
\defcitealias{pilyugin&grebel16}{Scal-PG16}
\defcitealias{marino13}{O3N2-M13}
\defcitealias{curti20}{O3N2-C20}
\defcitealias{wang&lilly21}{WL21}
\newcommand{\calzetti}{\citetalias{calzetti00}}
\newcommand{\chabrier}{\citetalias{chabrier03}}
\newcommand{\bpt}{\citetalias{baldwin81}}
\newcommand{\dopita}{\citetalias{dopita16}}
\newcommand{\pg}{\citetalias{pilyugin&grebel16}}
\newcommand{\marino}{\citetalias{marino13}}
\newcommand{\curti}{\citetalias{curti20}}
\newcommand{\wl}{\citetalias{wang&lilly21}}

\usepackage{titlesec}
\titleclass{\subsubsubsection}{straight}[\subsection]
\newcounter{subsubsubsection}[subsubsection]
\renewcommand\thesubsubsubsection{\thesubsubsection.\arabic{subsubsubsection}}
\titleformat{\subsubsubsection}
  {\normalfont\normalsize\bfseries}{\thesubsubsubsection}{1em}{}
\titlespacing*{\subsubsubsection}
{0pt}{3.25ex plus 1ex minus .2ex}{1.5ex plus .2ex}

\newcommand{\aref}[1]{\hyperref[#1]{Appendix~\ref{#1}}}

\title[$Z_\mathrm{gas}$--$\Sigma_\mathrm{SFR}$ relation]{MAUVE--MUSE: When Metallicity Follows or Fights Star Formation---A Mass-Dependent Inversion in Virgo Galaxies}

\author[R. Huang et al.]{Rongjun Huang (\CJKname)\orcidlink{0000-0002-6646-8365}$^{1,2}$\thanks{E-mail: Rongjun.Huang@icrar.org, Astro@Rongjun-Huang.com} 
Luca Cortese\orcidlink{0000-0002-7422-9823}$^{1}$, 
Barbara Catinella\orcidlink{0000-0002-7625-562X}$^{1}$, 
Luke J. M. Davies\orcidlink{0000-0003-3085-0922}$^{1}$, 
\newauthor
Toby Brown\orcidlink{0000-0003-1845-0934}$^{2}$, 
Andrei Ristea\orcidlink{0000-0003-2723-0810}$^{3,4}$, 
Alessandro Boselli\orcidlink{0000-0002-9795-6433}$^{5}$, 
Andrew J. Battisti\orcidlink{0000-0003-4569-2285}$^{1,6}$, 
Vicente Villanueva\orcidlink{0000-0002-5877-379X}$^{7,8}$, 
\newauthor
Kristine Spekkens\orcidlink{0000-0002-0956-7949}$^{9}$, 
Sara L. Ellison\orcidlink{0000-0002-1768-1899}$^{10}$,
Daniel A. Dale\orcidlink{0000-0002-5782-9093}$^{11}$,  
Sabine Thater\orcidlink{0000-0003-1820-2041}$^{12}$, 
Amirnezam Amiri\orcidlink{0000-0002-8553-1964}$^{13,14}$
\\
$^{1}$International Centre for Radio Astronomy Research, University of Western Australia, 7 Fairway, Crawley, 6009, Western Australia, Australia\\ 
$^{2}$National Research Council of Canada, Herzberg Astronomy and Astrophysics Research Centre, 5071 W. Saanich Rd., Victoria, BC, V9E 2E7, Canada\\
$^{3}$Centre for Astrophysics and Supercomputing, Swinburne University of Technology, John St, Hawthorn, 3122, Victoria, Australia\\
$^{4}$ARC Centre of Excellence in Optical Microcombs for Breakthrough Science (COMBS), Australia\\
$^{5}$Aix-Marseille Université, CNRS, CNES, LAM, Marseille, France\\
$^{6}$Research School of Astronomy and Astrophysics, Australian National University, Cotter Road, Weston Creek, ACT, 2611, Australia\\
$^{7}$Instituto de Estudios Astrofísicos, Facultad de Ingeniería y Ciencias, Universidad Diego Portales, Av. Ejército Libertador 441, 8370191 Santiago, Chile\\
$^{8}$Millennium Nucleus for Galaxies, MINGAL\\
$^{9}$Department of Physics, Engineering Physics and Astronomy, Queen's University, Kingston, ON, K7L 3N6, Canada\\
$^{10}$Department of Physics \& Astronomy, University of Victoria, Finnerty Road, Victoria, BC, V8P 1A1, Canada\\
$^{11}$Department of Physics \& Astronomy, University of Wyoming, Laramie, WY 82071, USA\\
$^{12}$Department of Astrophysics, University of Vienna, T¨urkenschanzstrasse
17, 1180 Wien, Austria\\
$^{13}$School of Astronomy, Institute for Research in Fundamental Sciences (IPM), Tehran, P.O. Box 19395-5531, Iran\\
$^{14}$Department of Physics, University of Arkansas, 226 Physics Building, 825 West Dickson Street, Fayetteville, AR 72701, USA\\
}

\pubyear{\the\year{}}

\begin{document}
\label{firstpage}
\pagerange{\pageref{firstpage}--\pageref{lastpage}}
\maketitle

\begin{abstract}
Although globally-integrated studies often find that, at fixed stellar mass, high star formation rate (SFR) galaxies are relatively metal-poor while lower-SFR systems are more metal-rich, the corresponding coupling between gas-phase metallicity ($Z_{\rm gas}$) and star formation on sub-galactic scales remains poorly constrained. In this study, we analyse 14 Virgo spiral galaxies from the MAUVE--MUSE survey to revisit the resolved mass--metallicity relation (rMZR) and its secondary dependence on SFR surface density ($\Sigma_\mathrm{SFR}$) at $\sim 100$\,pc scales. We construct co-spatial maps of stellar mass surface density ($\Sigma_*$), $\Sigma_\mathrm{SFR}$ and gas-phase oxygen abundance. MAUVE--MUSE galaxies follow a standard rMZR, but once binned in $\Sigma_*$, we find a clear mass-dependent inversion in the $Z_\mathrm{gas}$--$\Sigma_\mathrm{SFR}$ relation with O3N2-based metallicity calibrations: the commonly reported anti-correlation is confined to low-$\Sigma_*$ bins, whereas high-$\Sigma_*$ regions show a positive correlation, with an inversion point at $\log_{10}(\Sigma_*/M_\odot\,\mathrm{kpc}^{-2})\simeq 7.5$--8.0. Both correlated and anti-correlated \ion{H}{ii} regions can coexist within the same discs, and the observed mass dependence emerges only when grouping \ion{H}{ii} spaxels by $\Sigma_*$. We develop a spatially resolved gas--regulator model and show that the observed correlation and anti-correlation between $Z_\mathrm{gas}$ and $\Sigma_\mathrm{SFR}$ arise from the competition between star-formation-driven and gas-supply-driven variability: the sign of the local $Z_\mathrm{gas}$--$\Sigma_\mathrm{SFR}$ relation is set by which of these dominates. This framework can be naturally extrapolated to the integrated scenario, providing a unified explanation for both the resolved and global $Z_\mathrm{gas}$--$\mathrm{SFR}$ relations. However, the presence and strength of the $Z_\mathrm{gas}$--$\Sigma_\mathrm{SFR}$ (anti-)correlation depend strongly on the metallicity indicator used, highlighting the challenge of disentangling physical secondary trends within metallicity scaling relations.
\end{abstract}




\section{Introduction}
\label{sec:intro}

The global stellar mass--gas-phase metallicity relation \citep[MZR; e.g.,][]{lequeux79,tremonti04,erb06, maiolino08, zahid14, curti20, sanders21}, demonstrates that gas-phase metallicity ($Z_{\rm gas}$, also commonly referred to as gas-phase oxygen abundance) increases with galaxy stellar mass ($M_*$), such that more massive galaxies host a more metal-enriched interstellar medium (ISM). Physically, this trend is commonly interpreted as the outcome of mass-dependent chemical regulation: in the low-mass regime, the near-linear trend is a consequence of the integrated star formation history, whereas the high-mass flattening is attributed to stronger regulation by outflows and/or saturation toward the maximum oxygen yield (see \citealt{maiolino&mannucci19} for a review). While this global relation provides an essential benchmark for obtaining a comprehensive picture of chemical evolution studies, the advent of integral-field spectroscopy (IFS) has shifted attention toward understanding how metallicity is regulated within galaxies on a resolved basis. By resolving discs into individual star-forming spaxels, IFS surveys now enable direct measurement of local scaling relations between gas-phase oxygen abundance, and local stellar and star formation properties (see \citealt{sanchez20,sanchez21a} for a review).

The Calar Alto Legacy Integral Field Area \citep[CALIFA;][]{sanchez13,sanchez17,cresci19} survey revealed that the global MZR has a local counterpart: on kpc-scale spatial resolution, gas-phase oxygen abundance correlates tightly with the local stellar mass surface density ($\Sigma_*$). This resolved mass--metallicity relation (rMZR) was subsequently confirmed using the much larger and more diverse samples including the Mapping Nearby Galaxies at APO \citep[MaNGA;][]{barreraballesteros16,baker23} and the Sydney-AAO Multi-object Integral-field spectrograph \citep[SAMI;][]{sanchez19} surveys, establishing the rMZR as a ubiquitous feature of star-forming discs on kpc scales in the local Universe. More recently, IFS surveys based on Very Large Telescope/Multi-Unit Spectroscopic Explorer (VLT/MUSE) have extended this resolved scaling relation to intermediate redshift ($0.1\lesssim z \lesssim 1$) while retaining comparable spatial resolution ($\sim$kpc scale). In particular, the MUSE--Wide survey demonstrated the existence of an rMZR at $z\sim0.26$ \citep{yao22}, with a shape similar to that at $z\sim0$ but a modestly lower normalization, consistent with our models of chemical evolution over cosmic time. Likewise, the Middle Ages Galaxy Properties with Integral Field Spectroscopy \citep[MAGPI;][]{foster21} survey confirmed an rMZR in galaxies at $z\sim0.3$ on a similar (kpc-scale) spatial resolution \citep{koller24}. Additionally, with the James Webb Space Telescope/Near Infrared Spectrograph (JWST/NIRSpec), the ALPINE–CRISTAL–JWST programme is now extending rMZR studies to representative main-sequence galaxies at $z\simeq 4$--6 at resolutions of a few hundred pc, with $Z_\mathrm{gas}$ showing a stronger dependence on $\Sigma_{\rm SFR}$ than those in local universe \citep{fujimoto25}. Taken together, these IFS studies indicate that gas-phase oxygen abundance correlates more tightly with local stellar mass surface density than with galactocentric radius alone \citep[e.g.,][]{barreraballesteros16,baker23}: at fixed radius, spaxels with higher $\Sigma_*$ tend to be more metal-rich, and a similar $\Sigma_*$–$Z_{\rm gas}$ relation is recovered in galaxies with very different sizes and radial profiles. In this sense, the rMZR appears to reflect local physics, even though it is closely linked to the underlying radial structure of galactic discs.

On global scales, the relation between gas-phase metallicity, stellar mass, and star formation rate (SFR) remains actively debated. While several studies have interpreted the classical MZR as a projection of a more fundamental $M_*$--$Z_{\rm gas}$--SFR surface/plane, often referred to as the Fundamental Metallicity Relation \citep[FMR; e.g.,][]{ellison08, mannucci10, laralopez10, yates12, zahid13, bothwell13, peng14, bothwell16, gao18, cresci19, curti20, pallottini25}, other works have found that the secondary dependence of metallicity on SFR is weaker than originally proposed, not always detected, or sensitive to sample selection and abundance calibration \citep[e.g.,][]{sanchez13,sanchez17,sanchez19,barreraballesteros17}. In most studies that do recover an FMR-like trend, galaxies with higher SFR at fixed $M_*$ tend to be more metal-poor, i.e.\ $Z_{\rm gas}$ and SFR are anti-correlated, a behaviour often interpreted as the signature of metal-poor gas accretion diluting the ISM and fuelling star formation. This broader uncertainty naturally raises the question of whether an analogous three-parameter relation exists on resolved scales, i.e.\ whether the rMZR carries a meaningful secondary dependence on local SFR surface density ($\Sigma_{\rm SFR}$). Several works based on large, predominantly star-forming disc samples report that once the primary $\Sigma_*$ dependence is removed, residual trends with $\Sigma_{\rm SFR}$ are weak or absent on kpc physical scales, suggesting that local metallicity primarily traces longer-term mass assembly and recycling \citep[e.g.,][]{sanchez13,yao22}. Similarly, some cosmological simulations also predict that the spatially resolved analogue of the FMR (i.e. rFMR) on $\sim$kpc scales is weak \citeg{qi25}. However, recent studies find a statistically significant $Z_{\rm gas}$--$\Sigma_{\rm SFR}$ (anti-)correlation and motivate a rFMR, wherein metallicity at fixed $\Sigma_*$ depends on $\Sigma_{\rm SFR}$ \citep{sanchezmenguiano19,baker23,koller24}. Particularly, the MAGPI data present a clear signal that the \emph{sign} of this correlation can reverse with $\Sigma_*$: low-$\Sigma_*$ regions show an anti-correlation between $Z_{\rm gas}$ and $\Sigma_{\rm SFR}$, while high-$\Sigma_*$ regions exhibit a flat or positive relation \citep{koller24}. Global and resolved studies therefore suggest that any FMR- or rFMR-like behaviour may not be universal, but instead depend on physical regime, sample selection, and metallicity calibration.

The potential flip between the $Z_{\rm gas}$--$\Sigma_{\rm SFR}$ correlation and anti-correlation suggests that metallicity variations are not simply residual scatter around the rMZR, but may encode the local gas--star-formation cycle. This view is consistent with recent work treating gas-phase metallicity residuals as spatially correlated fluctuation fields generated by stochastic enrichment and smoothed by turbulent mixing \citeg{krumholz&ting18,krumholz25,li25,zhang26}. While such metallicity-fluctuation frameworks provide a useful description of the spatial coherence of abundance residuals, interpreting the \emph{sign} of the local $Z_{\rm gas}$--$\Sigma_{\rm SFR}$ coupling requires a framework that explicitly connects metal enrichment, dilution, and star formation. A widely used description of this link is provided by the ``gas--regulator'' or ``bathtub'' models, in which the gas content and metal budget of a galaxy (or region of a galaxy) are governed by the balance between gas inflow, star formation, and outflows \citep[e.g.,][]{tinsley76,dave12,lilly13,peng14,forbes14,belfiore19,sharda21}. In their galaxy-integrated form, these models predict that the gas-phase metallicity tends toward an effective equilibrium set by the metal yield, the metallicity of the accreted gas, and the interplay between star formation and inflow rate, naturally reproducing both the shape of the global MZR and its commonly observed anti-correlation with SFR. A more explicit connection between time-dependent star formation and metallicity has been established by \citet[][hereafter \wl]{wang&lilly21}, who treated individual ``gas–regulator units'' (from galaxy-wide down to $\sim 100$\,pc scales) as reservoirs subject to time-variable inflow rate and star formation efficiency (SFE). Within the \wl\ framework, the residual correlation between metallicity and SFR at fixed stellar mass (or gravitational potential) is set by which source of variability dominates: when fluctuations in SFE drive changes in the SFR at roughly fixed inflow rate, metallicity and SFR are predicted to have a positive correlation, whereas strongly time-variable inflow at nearly constant SFE produces metal dilution and thus a negative correlation between metallicity and SFR. However, testing such analytical frameworks on local scales requires spatially resolved data that simultaneously (i) resolve individual (or clumps of) \ion{H}{ii} regions on $\sim 100$\,pc scales, comparable to the physical resolution of the MUSE Atlas of Disks (MAD) survey \citep{denbrok20} and the Physics at High Angular Resolution in Nearby Galaxies (PHANGS)--MUSE \citep{emsellem22} survey; and (ii) span a broad dynamic range in $\Sigma_*$, so that both SFR-driven and inflow-driven regimes can be probed within and across galaxies.

In this work, we use observations of the first 14 Virgo Cluster spirals observed as part of the Multiphase Astrophysics to Unveil the Virgo Environment with MUSE (MAUVE--MUSE) survey \citep{mauve} to revisit the rMZR and its secondary dependence on $\Sigma_{\rm SFR}$ at $\sim 100$\,pc resolution. We show that the Virgo cluster spirals sample exhibits a robust mass-dependent inversion in both the $Z_{\rm gas}$--$\Sigma_{\rm SFR}$ relation and in the corresponding residual space ($\Delta Z_{\rm gas}$--$\Delta \Sigma_{\rm SFR}$ relation), where $\Delta Z_{\rm gas}$ and $\Delta \Sigma_{\rm SFR}$ denote offsets at fixed $\Sigma_*$ from the mean resolved trends. This behaviour is consistent with previous indications of a sign reversal, but extending them to a cluster sample at $\sim 100$\,pc resolution for the first time. To characterise the spatial heterogeneity of this coupling, we introduce a local bivariate Moran-like statistic \citep{moran50,anselin95} that quantifies, around each spaxel, whether nearby regions tend to show same-sign or opposite-sign pairs of $(\Delta Z_{\rm gas},\,\Delta\Sigma_{\rm SFR})$, thereby identifying spatially coherent zones as concordant or discordant. Finally, we develop a spatially resolved regulator model to interpret these trends in our data. The results demonstrate that the observed mass-dependent sign flip in the $Z_{\rm gas}$--$\Sigma_{\rm SFR}$ relation is governed by the relative amplitude of local star-formation versus gas-supply variability. 

This paper is organized as follows: \autoref{sec:data-methods} introduces the data and methods used in this study, \autoref{sec:results} outlines our results, \autoref{sec:origin} presents the spatially resolved framework, its interpretation and the extrapolation to a global integrated scenario, \autoref{sec:discussions} discusses the robustness of our findings across different metallicity calibrations, star-forming spaxel selections, and the special case of the metallicity outlier NGC~4383, and \autoref{sec:conclusions} outlines our conclusions. Throughout this work, we adopt a distance of 16.5 Mpc for Virgo Cluster members \citep{mei07}, aligning with previous and ongoing MAUVE studies; the flat Lambda Cold Dark Matter ($\Lambda$CDM) model is adopted by assuming the Hubble constant is $H_0$ = 70\,$\mathrm{kms^{-1}Mpc^{-1}}$ and the mass density of the Universe is $\Omega_{\rm{m,0}}$ = 0.3.

\section{Data \& Methods}
\label{sec:data-methods}

\subsection{MAUVE--MUSE Data}
\label{sec:mauve-products}

The data analysed in this work are drawn from the internal release~v2 (December~2024) of the MAUVE--MUSE survey. Designed to investigate how the cluster environment shapes the gas–star formation cycle, MAUVE targets 40 late-type Virgo Cluster galaxies using a suite of multi-wavelength facilities \citep{mauve}. For each target, we use the fully reduced MUSE datacubes produced with \texttt{pymusepipe}~v2.28.2 and \texttt{esorex}~3.13.6, following closely the reduction, sky subtraction, and flux-calibration procedures described by \citet{emsellem22}. The data for each galaxy is delivered as a science-ready, fully-reduced FITS file on the native 0\farcs2 pixel grid over the full $4750$--$9350$ \AA\ spectral range of VLT/MUSE. 

Stellar-continuum and emission-line maps used in this paper are derived from these cubes using the \texttt{nGIST} pipeline \citep{bittner19,fraser-mckelvie25}, a \texttt{pPXF}-based \citep{cappellari04, cappellari16, cappellari23} wrapper optimised for MAUVE--MUSE data. Briefly, we first construct an adaptively binned representation of each cube by applying the Voronoi algorithm of \citet{cappellari04} to the stellar continuum in the $4800$--$7000$\,\AA\ spectral window; spaxels are then aggregated until the bin-level $\mathrm{S/N}$, computed by summing signal and adding noise in quadrature, reaches a target of $\mathrm{S/N}\ge40$. The resulting Voronoi tessellation is then adopted as the common spatial grid for all \texttt{nGIST} products. This approach ensures strict one-to-one correspondence between all the following derived properties without any additional resampling or smoothing. The MAUVE--MUSE observations have a typical seeing full width at half maximum (FWHM) of $\sim 1''$, corresponding to a physical scale of $\sim 80$\,pc at the Virgo Cluster's distance, so that the effective angular resolution is set by the point spread function (PSF) rather than by the MUSE sampling pixel scale of 0\farcs2\,pixel$^{-1}$. The $\mathrm{S/N}\ge40$ Voronoi binning on the stellar continuum typically groups a few up to several tens of spaxels per bin with 95\% containing fewer than 40 spaxels, yielding characteristic bin sizes comparable to the seeing footprint and hence also of order $\sim 100$\,pc. Throughout this work, we therefore refer to our spatially resolved measurements as probing $\sim 100$~pc scales.

On this Voronoi grid of each galaxy, the stellar continuum and kinematics in each bin were fit with \texttt{pPXF} using \textsc{MILES}\footnote{MILES website: \url{https://research.iac.es/proyecto/miles/}} Simple Stellar Population (SSP) templates \citep{vazdekis10}, after convolving the templates with a wavelength-dependent Gaussian MUSE line–spread function (LSF, parameterised following Equation 8 of \citealt{bacon17}; see also \citealt{emsellem22}). Multiplicative Legendre polynomials are included to account for residual relative flux calibration offsets. After subtracting the best-fitting stellar model, Gaussian profiles are fitted simultaneously to the main nebular emission lines (e.g.\ H$\alpha$, H$\beta$, [\ion{O}{iii}], [\ion{N}{ii}], [\ion{S}{ii}]), yielding maps of observed line fluxes, velocities, and velocity dispersions with associated uncertainties.

While MAUVE--MUSE observations are still in progress, a total of 14 galaxies have been fully observed, reduced and therefore analyzed in this paper (see \citealt{mauve} for full details of the MAUVE--MUSE sample): IC~3392, NGC~4064, NGC~4192, NGC~4293, NGC~4298, NGC~4330, NGC~4383, NGC~4396, NGC~4419, NGC~4457, NGC~4501, NGC~4522, NGC~4694, and NGC~4698.

\subsection{Stellar Mass Surface Density}
\label{sec:stellar-mass}

We first derive spatially resolved stellar mass surface densities from the MAUVE--MUSE datacubes and the associated \texttt{nGIST} value-added products. For each galaxy, we start from the \texttt{nGIST} star-formation-history (SFH) products, which provide, for every Voronoi bin, a vector of SSP template weights (i.e. the relative light contributions of each stellar metallicity-age component in the spectral fit) and the corresponding SSP parameters grid in stellar metallicity and age, $([{\rm M/H}], \log_{10}(t))$. We then adopt the \textsc{MILES}-based SSP \citep{vazdekis10} predictions based on \textsc{BaSTI} isochrone grids \citep{pietrinferni04} with a \citet[][hereafter \chabrier]{chabrier03} initial mass function (IMF), using their tabulated Johnson $R$-band mass-to-light ratios ($(M/L)_R$) as a function of $([{\rm M/H}], \log_{10} t)$. We use the $R$ band as the \textsc{MILES} SSP templates provide the $(M/L)$ in Johnson bands, and the $R$ band is fully and comfortably covered by MUSE, reducing band-edge systematics relative to $V$ and $I$ bands. The SFH weights for each Voronoi bin are then combined with this $(M/L)_R$ grid to compute a light-weighted $R$-band mass-to-light ratio,
\begin{equation}
    (M/L)_R^{\rm bin} = \sum_k w_k \,(M/L)_{R,k},
\end{equation}
where $w_k$ and $(M/L)_{R,k}$ are the SFH weight and $R$-band mass-to-light ratio of the $k$-th SSP component in the bin based on variable stellar metallicities and ages. This procedure yields a resolved map of $(M/L)_R^{\rm bin}$ defined over all Voronoi zones.

To obtain the corresponding $R$-band luminosity per bin, we construct the observed spectral energy distribution, $F_\lambda$, and convolve it with the \texttt{bessell-R} filter response using the \texttt{speclite} package \citep{kirkby23}. This yields an AB $R$-band magnitude, $m_R$, and flux (in nanomaggies) per spaxel. We then collapse these spaxel-level fluxes onto the Voronoi tessellation by averaging, only over spaxels assigned to that bin, while preserving the survey's native masking of invalid pixels. The resulting bin-averaged AB magnitudes are corrected for Galactic extinction using the \texttt{nGIST} $E(B-V)$ map and a \citet[][hereafter \calzetti]{calzetti00} attenuation curve. These SFH weights are obtained from the full \texttt{nGIST}/\texttt{pPXF} spectral fit, following the same general approach adopted in PHANGS \citep{pessa21}, where the internal stellar attenuation is already accounted for within the fit. Corrected apparent magnitudes ($m_R^{\rm corr}$) are converted into absolute magnitudes via the distance modulus corresponding to the adopted Virgo Cluster distance of 16.5 Mpc \citep{mei07}, and hence into $R$-band luminosities ($L_R$) in solar units using the solar $R$-band absolute magnitude, $M_{R,\odot}=4.61$ \citep{willmer18}. 

The stellar mass in each Voronoi bin is then obtained as
\begin{equation}
    M_*^{\rm bin} = L_R^{\rm bin} \times (M/L)_R^{\rm bin}.
\end{equation} 
Dividing $M_*^{\rm bin}$ by the corresponding projected bin area yields a map of projected stellar mass surface density, $\Sigma_*$ (in $M_\odot\,{\rm kpc}^{-2}$). 

Finally, we correct observed surface densities for disc inclination, adopting inclination angles ($i$) from the Virgo Environment Traced in CO \citep[VERTICO;][]{brown21} survey, where $i=0^\circ$ is face-on and $b/a$ is the axis ratio in projection. Assuming an oblate disc geometry with intrinsic thickness, $q_0 = 0.2$ \citep{holmberg58,tully09,cortese16}, we compute the inclination correction factor for each galaxy: 
\begin{equation}
    \frac{b}{a} = \sqrt{(1-q_0^2)\cos^2 i + q_0^2},
\label{eq:inclination}
\end{equation}
which encodes the observed flattening of a disc of finite thickness. The $\Sigma_*$ and the $\Sigma_\mathrm{SFR}$ in the following \autoref{sec:sfr} are then multiplied by this $b/a$ factor to correct from the projected to the intrinsic face-on surface densities (i.e. to remove foreshortening). This correction addresses geometry only and does not explicitly model inclination-dependent attenuation. 

While all spectral analyses (including the stellar population fitting here and the nebular emission fitting below) are performed on the Voronoi-binned spectra to ensure a minimum continuum $\mathrm{S/N} \geq 40$, we reassign the derived properties of each bin back to its constituent spaxels to create resolved maps and undertake statistical analysis, following the standard ``dezonification'' procedure introduced by \citet{cidfernandes13}. This procedure preserves the survey's native spatial sampling of $0\farcs2\,\mathrm{pixel}^{-1}$, effectively weighting each Voronoi bin by its projected area. Consequently, unless otherwise noted, the term ``spaxel'' in the following contexts refers to these grid elements on the native scale, which carry the robust physical values derived from their parent Voronoi bin. Additionally, in agreement with the results of \citet{koller24}, we found that adopting a bin-based analysis does not significantly change the results presented in this work compared to the current spaxel-based approach.

\subsection{Star Formation Rate Surface Density}
\label{sec:sfr}

Nebular dust attenuation is quantified from the Balmer Decrement (${\rm BD}$) measured on the same Voronoi grid as all other MAUVE--MUSE products. For each spaxel, we start from the \texttt{nGIST} H$\alpha$ and H$\beta$ flux and error maps ($F_{\rm line}$ and $\sigma_{\rm line}$, where $\mathrm{line}=\mathrm{H}\alpha$ and H$\beta$) and impose a basic quality-control cut requiring $F_{\rm line}/\sigma_{\rm line} \geq 3$, together with a minimum flux threshold of $F_{\rm line} \geq 20\times10^{-20}\,{\rm erg\,s^{-1}\,cm^{-2}}$ per spaxel in the native MAUVE--MUSE units. Spaxels that do not meet this criteria turn out to be spaxels that do not full fill our additional criteria to isolate star-forming spaxels (see description below). As such, even relaxing such a quality cut would not impact our final selection of star-forming spaxels. The Balmer Decrement is defined as ${\rm BD} = F_{\mathrm{H}\alpha}/F_{\mathrm{H}\beta}$ (using observed fluxes) and we impose a floor corresponding to the Case~B recombination value at ${\rm BD}=2.86$, appropriate for electron temperature and number density at $T_{\rm e} = 10^4$ K and $n_{\rm e} = 100\,{\rm cm^{-3}}$\footnote{We adopt this standard Case-B value as a fixed reference throughout, noting that modest variations with nebular temperature and density are possible but are expected to contribute only a secondary systematic uncertainty compared with the larger uncertainties associated with line-flux measurements and calibration choice \citeg{congiu23}.} \citep{osterbrock&ferland06}, i.e.\ any measured ${\rm BD}<2.86$ is set to 2.86 to avoid unphysical negative colour excesses. The corresponding colour excess is obtained using the following form \citep{osterbrock&ferland06}:
\begin{equation}
    E(B-V)_{\rm gas}
    = \frac{2.5}{k(\mathrm{H}\beta) - k(\mathrm{H}\alpha)}
      \log_{10}\!\left(\frac{{\rm BD}}{2.86}\right),
\end{equation}
where $k(\lambda)$ is the Milky Way-like extinction curve with the effective obscuration at $V$-band $R_V = 3.1$ \citep{cardelli89}. 
All strong nebular lines entering the subsequent analysis are then dereddened with this $E(B-V)_{\rm gas}$ via
\begin{equation}
    F_{\lambda}^{\rm corr}
    = F_{\lambda}\,10^{0.4\,E(B-V)_{\rm gas}\,k(\lambda)}.
\end{equation}

\begin{figure*}
    \centering
    \includegraphics[width=\textwidth]{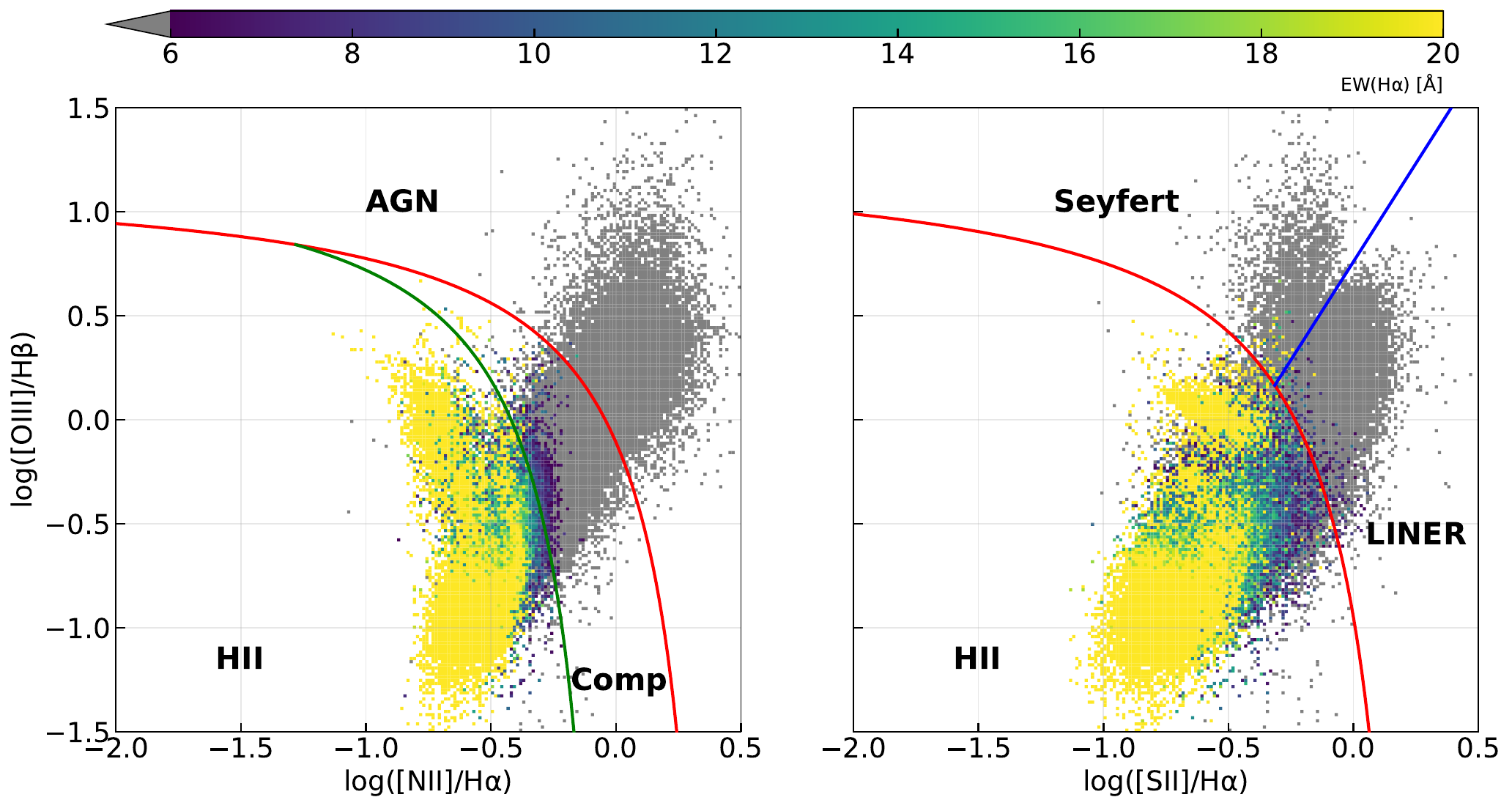}
    \caption{[\ion{N}{ii}] (left) and [\ion{S}{ii}] (right) \bpt\ diagrams for ``classified'' spaxels of all 14 MAUVE--MUSE galaxies. Data points correspond to Voronoi spaxels classified using the demarcation lines within $1\sigma$ confidence described in \autoref{sec:sfr}, colour-coded by $\mathrm{EW}(\mathrm{H}\alpha)$. Red and green curves show the \citet{kewley01} and \citet{kauffmann03a} demarcation lines in the [\ion{N}{ii}] diagram, while red and blue curves indicate the \citet{kewley01} and \citet{kewley06} divisions in the [\ion{S}{ii}] diagram. Spaxels classified as star-forming (\textbf{HII} spaxels) require identification as \ion{H}{ii} regions on both \bpt\ diagrams, with $\mathrm{EW}(\mathrm{H}\alpha) > 6$\,\AA\ and $\sigma(\mathrm{H}\alpha) < 45\,\mathrm{km\,s^{-1}}$ cuts.}
    \label{fig:bpt}
\end{figure*}

Using these dust-corrected fluxes we construct the standard [\ion{N}{ii}] and [\ion{S}{ii}] \citet[][hereafter \bpt]{baldwin81} ratios, namely $F_{[\ion{N}{ii}]\lambda6583}/F_{\mathrm{H}\alpha}$, $F_{[\ion{S}{ii}](\lambda6716+\lambda6731)}/F_{\mathrm{H}\alpha}$, and $F_{[\ion{O}{iii}]\lambda5007}/F_{\mathrm{H}\beta}$, propagating flux-to-error ratios into the corresponding line-ratio uncertainties (see \autoref{fig:bpt}). Each spaxel is classified on both the [\ion{N}{ii}] and [\ion{S}{ii}] \bpt\ diagrams according to the \citet{kewley06} demarcation curves, and we require that the central value and its $\pm1\sigma$ error bars $(\log_{10}({[\ion{O}{iii}]/H\beta}),\log_{10}({\rm line/H\alpha}))$ all remain within the same \bpt\ region. Only spaxels that satisfy this criterion and the quality-control cuts in all lines entering a given diagram are assigned a secure class ($6\%$ spaxels that straddle a demarcation curve at the $\pm1\sigma$ level are excluded).

We define star-forming spaxels (hereafter \textbf{HII} mask) according to three criteria: (i) those that lie in the \ion{H}{ii} regime of both [\ion{N}{ii}] and [\ion{S}{ii}] \bpt\ diagrams and whose $\pm1\sigma$ error bars also remain entirely within the \ion{H}{ii} regime on both diagrams, with robust detections in all lines entering the classification; (ii) the equivalent width (EW) of H$\alpha$, $\mathrm{EW(H}\alpha) > 6$\,\AA\ \citep{cidfernandes10,lacerda18,lacerda20}; and (iii) the intrinsic H$\alpha$ velocity dispersion (after accounting for the MUSE instrument broadening), $\sigma(\mathrm{H}\alpha) < 45\,\mathrm{km\,s^{-1}}$ \citep{egorov23}, to exclude spaxels associated with extended outflow-dominated emission. All SFR and metallicity maps in our main analysis are therefore computed exclusively using this \textbf{HII} mask, and all subsequent resolved analyses implicitly refer to these \ion{H}{ii}-classified star-forming regions. Across the sample, the conservative \textbf{HII} selection retains 28\% of the total valid spaxels used in this work (with substantial galaxy-to-galaxy variation from 4\% to 52\%). Additionally, in \autoref{sec:discussion_sf}, we perform a robustness test by relaxing this conservative star-forming restriction and repeating the analysis with an expanded \textbf{HII+Comp} mask that also includes spaxels classified as composite on both \bpt\ diagrams, while still applying the same $\mathrm{EW(H}\alpha) > 6$\,\AA\ and $\sigma(\mathrm{H}\alpha) < 45\,\mathrm{km\,s^{-1}}$ cuts.

$\Sigma_\mathrm{SFR}$ maps are constructed from the dust-corrected H$\alpha$ flux. After applying the Balmer-decrement attenuation correction described above, we apply the \citet{calzetti07} calibration, rescaled to a \chabrier\ IMF by \citet{kennicutt&evans12}, to obtain 
\begin{equation}
    {\rm SFR}\,[M_\odot\,{\rm yr}^{-1}] =
    4.98 \times 10^{-42}\,L_{\mathrm{H}\alpha}^{\rm corr}\,[{\rm erg\,s^{-1}}].
\end{equation}
Similarly, we take the projected surface density of SFR and account for the inclination by \autoref{eq:inclination}, obtaining corrected $\Sigma_\mathrm{SFR}$. 

\subsection{Gas-phase Metallicity}
\label{sec:Z}

Gas-phase oxygen abundances\footnote{Throughout this paper, we use $[\mathrm{O/H}] \equiv 12+\log_{10}(\mathrm{O/H})$ to denote the gas-phase oxygen abundance, with a surface solar abundance of $12+\log_{10}(\mathrm{O/H})=8.69$ \citep{asplund09}.} are inferred from the dust-corrected emission-line maps, restricted to the \textbf{HII} spaxels defined in \autoref{sec:sfr}. Using only spaxels that meet the S/N and star-forming selections, we focus on a subset of four widely used strong-line prescriptions that are robustly applicable across the MAUVE--MUSE sample: the N2S2-H$\alpha$ calibration of \citet[][hereafter \dopita]{dopita16}, the S-calibration of \citet[][hereafter \pg]{pilyugin&grebel16}, the O3N2 index calibration of \citet[][hereafter \marino]{marino13}, and the O3N2 calibration of \citet[][hereafter \curti]{curti20}. The line ratios entering these prescriptions are listed in \autoref{tab:line-ratios}. Similar to \citet{watts24}, we reconstruct the total [\ion{N}{ii}]~$\lambda\lambda6548,6583$ and [\ion{O}{iii}]~$\lambda\lambda4959,5007$ fluxes by scaling the measured [\ion{N}{ii}]~$\lambda6583$ and [\ion{O}{iii}]~$\lambda5007$ fluxes by fixed factors of 1.34 and 1.33, respectively, given by theoretical line ratios in quantum mechanics \citep{storey&zeippen00}. All strong-line calibrations are applied within the recommended range $7.63 \leq [\mathrm{O/H}] \leq 9.23$ by \citet{kewley19}.

\begin{table}
    \caption{Definition of key emission-line ratios used in the extinction, SFR, and metallicity analysis, where all fluxes are dust-corrected. Here we adopt $F_{[\ion{N}{ii}](\lambda6548+\lambda6584)} = 1.34F_{[\ion{N}{ii}]\lambda6584}$ and $F_{[\ion{O}{iii}](\lambda4959+\lambda5007)} = 1.33F_{[\ion{O}{iii}]\lambda5007}$. }
    \centering
    \begin{tabular}{cc}
        \hline
        Line ratio & Definition \\
        \hline\hline
        $N2$    & $\displaystyle \frac{F_{[\ion{N}{ii}](\lambda6548+\lambda6584)}}{F_{\mathrm{H}\beta}}$ \\
        \hline
        $S2$    & $\displaystyle \frac{F_{[\ion{S}{ii}](\lambda6717+\lambda6731)}}{F_{\mathrm{H}\beta}}$ \\
        \hline
        $R3$    & $\displaystyle \frac{F_{[\ion{O}{iii}](\lambda4959+\lambda5007)}}{F_{\mathrm{H}\beta}}$ \\
        \hline
        $O3N2$  & $\displaystyle \log_{10}\!\left(
                      \frac{F_{[\ion{O}{iii}]\lambda5007}/F_{\mathrm{H}\beta}}
                          {F_{[\ion{N}{ii}]\lambda6583}/F_{\mathrm{H}\alpha}}
                   \right)$ \\
        \hline
        $N2S2$  & $\displaystyle \log_{10}\!\left(
                      \frac{F_{[\ion{N}{ii}]\lambda6583}}
                          {F_{[\ion{S}{ii}](\lambda6716+\lambda6731)}}
                   \right)$ \\
        \hline
        $y_\mathrm{D16}$   & $N2S2 + 0.264\,\log_{10}\!\left(
                      \frac{F_{[\ion{N}{ii}]\lambda6583}}
                          {F_{\mathrm{H}\alpha}}
                   \right)$ \\
        \hline

    \end{tabular}
    \label{tab:line-ratios}
\end{table}

The \dopita\ prescription combines the $N2S2$ and [\ion{N}{ii}]/H$\alpha$ ratios into a single abundance indicator. Using the shorthand in \autoref{tab:line-ratios}, we obtain the oxygen abundance as
\begin{equation}
    [\mathrm{O/H}]_{\rm D16}
    = 8.77 + y_\mathrm{D16} + 0.45(y_\mathrm{D16}+0.30)^5.
\end{equation}

The \pg\ calibration uses the $N2$, $S2$, and $R3$ ratios. We evaluate the branch-dependent polynomial expressions given in \citet{pilyugin&grebel16}:
\begin{equation}
    [\mathrm{O/H}]_{\rm PG16} =
    \begin{cases}
        8.424
        + 0.030\,\log_{10}(R3/S2)
        + 0.751\,\log_{10}(N2) \\[2pt]
        \qquad
        + \bigl[-0.349 + 0.182\,\log_{10}(R3/S2) \\[2pt]
        \qquad
        + 0.508\,\log_{10}(N2)\bigr]\log_{10}(S2),\  \\[2pt]
        \qquad
        \log_{10}(N2) \ge -0.6; \\[6pt]
        8.072
        + 0.789\,\log_{10}(R3/S2)
        + 0.726\,\log_{10}(N2) \\[2pt]
        \qquad
        + \bigl[1.069 - 0.170\,\log_{10}(R3/S2) \\[2pt]
        \qquad
        + 0.022\,\log_{10}(N2)\bigr]\log_{10}(S2),\  \\[2pt]
        \qquad
        \log_{10}(N2) < -0.6.
    \end{cases}
\end{equation}

For the \marino\ calibration, we adopt the O3N2 index defined in \autoref{tab:line-ratios} and use the corresponding abundance relation:
\begin{equation}
    [\mathrm{O/H}]_{\rm O3N2-M13}
    = 8.533 - 0.214\,O3N2,
\end{equation}

Finally, we also adopt the O3N2-based calibration of \citet{curti20}, which uses the same O3N2 index but fits a higher-order polynomial to stacked Sloan Digital Sky Survey \citep[SDSS;][]{abazajian09} spectra. We evaluate the recommended \curti\ functional form spaxel-by-spaxel over the valid range: 
\begin{equation}
    O3N2 = 0.281 - 4.765\,x - 2.268\,x^2,
\end{equation}
where $x \equiv [\mathrm{O/H}]_{\rm O3N2-C20} - 8.69$. In the course of this work, we also implemented the multi-indicator (``combined'') strategy advocated by \citet{curti20}. In this approach, the oxygen abundance is inferred for each spaxel by jointly comparing the observed set of available strong-line ratios ($R3$, $N2$, $S2$, $O3N2$ and $O3S2$; see Table 1 of \citealt{curti20}) to the corresponding empirical calibrations and selecting the best-fit solution by minimizing the $\chi^2$ statistics that accounts for the measurement uncertainties and the intrinsic calibration scatter. While this method yields self-consistent metallicity estimates with formally small uncertainties, we find that it can introduce diagnostic-driven discontinuities within individual galaxies when different ratios dominate in different regions. To ensure internal consistency and straightforward interpretation of spatial trends, we therefore adopt the single $O3N2$ calibration of \citet{curti20}. 

All metallicity prescriptions are evaluated exclusively on the conservative \textbf{HII} star-forming mask, on the same Voronoi grid as the stellar-mass and SFR surface-density maps. This guarantees that $\Sigma_*$, $\Sigma_{\rm SFR}$, and $12+\log_{10}({\rm O/H})$ are strictly co-spatial for every bin used in the resolved mass–metallicity–SFR analysis presented in \autoref{sec:results}. 
Throughout this paper, to be consistent, we adopt \marino\ as the standard prescription. We select this indicator to ensure robustness and comparability across studies, due to the following reasons. First, as demonstrated by \citet{mcleod21}, composite theoretical calibrations (e.g., \dopita, which combines $N2$ and [S {\sc ii}]/H$\alpha$) introduce non-negligible scatter compared to other empirical relations. Second, resolved metallicity maps are often susceptible to contamination from Diffuse Ionized Gas (DIG); \citet{kumari19} demonstrated that O3N2 is remarkably insensitive to DIG contamination (showing offsets of only $\sim$0.02 dex between \ion{H}{ii} and DIG regions), whereas indicators utilizing \ion{S}{ii} (such as \dopita\ and \pg) are heavily affected, a finding we also confirm in \autoref{sec:discussion_metallicity}. Third, both the \marino\ and \curti\ calibrations are derived from large IFS datasets (CALIFA and MaNGA, respectively), making them particularly well-suited for spatially resolved studies like ours. Finally, \marino\ is extensively employed in IFS studies investigating the rMZR \citeg{sanchez13,barreraballesteros16,errozferrer19,yao22,koller24}, ensuring our results are consistently comparable to the broader literature. However, we acknowledge the different results caused by different calibrations, and we will therefore discuss this effect in \autoref{sec:discussion_metallicity}. 

\subsection{Local Moran-like Correlation Index ($I_i$)}
\label{sec:moran}

To quantify how two spatially resolved fields co-vary on local scales, we construct a ``local Moran-like” product that combines the value of one property at a given position with the neighbourhood behaviour of a second property. This construction is inspired by the classical Moran's $I$ statistic for spatial autocorrelation \citep{moran50} and its generalization: local indicators of spatial association \citep{anselin95}. Such decomposing statistics were first introduced for astronomical use in \citet{arnold&wright24} to study the kinematic substructure of star clusters. Consider two generic maps $A(i)$ and $B(i)$ defined on the same set of valid spaxels $i$ (in this work, $A \equiv \mathrm{[O/H]}$ and $B \equiv \Sigma_{\rm SFR}$). We first compute the mean and standard deviation of each field over all valid locations, denoted $(\mu_A,\sigma_A)$ and $(\mu_B,\sigma_B)$, and form standardised residuals
\begin{equation}
    \begin{cases}
        z_A(i) = \frac{A(i)-\mu_A}{\sigma_A}, \\
        z_B(i) = \frac{B(i)-\mu_B}{\sigma_B}.
    \end{cases}
\end{equation}
The offsets $z_A(i)$ and $z_B(i)$ measure the local departure of each spaxel from the global behaviour of the sample in units of the intrinsic scatter, after subtracting the azimuthally averaged radial profile (or the other secondary dependence) of each field. Positive values indicate that a given location lies above the mean in that field, while negative values indicate a deficit relative to the mean.

To embed this information in its spatial context, we next characterise the local environment of $B$ around each spaxel $i$ by averaging the $z_B$ values of its neighbouring spaxels. In the present work, we adopt a simple eight-connected neighbourhood, so that the neighbour set $N(i)$ consists of the immediately adjacent spaxels in the two-dimensional grid and $N_i=8$ for interior pixels. The local ``spatial lag'' of $B$ at position $i$ is then defined as
\begin{equation}
    z_{\mathrm{lag},B}(i)
    \equiv \frac{1}{N_i}\sum_{j\in N(i)} z_B(j).
\end{equation}
This quantity is a dimensionless measure of whether the environment surrounding $i$ is systematically enhanced or suppressed in $B$ relative to the global mean, again in units of the intrinsic dispersion: $z_{\mathrm{lag},B}(i)>0$ implies that the neighbours tend to be above the mean in $B$, while $z_{\mathrm{lag},B}(i)<0$ indicates a locally depressed environment.

The final local Moran-like product at each position is then defined as
\begin{equation}
    I_i \equiv z_A(i)\,z_{\mathrm{lag},B}(i).
\end{equation}
This scalar encapsulates both the sign and the strength of the local correspondence between the value of $A$ at $i$ and the surrounding behaviour of $B$. When $z_A(i)$ and $z_{\mathrm{lag},B}(i)$ share the same sign, $I_i>0$ and the spaxel is ``in agreement'' with its environment: locations with $z_A(i)>0$ and $z_{\mathrm{lag},B}(i)>0$ correspond to regions that are elevated in $A$ and embedded in an environment that is likewise elevated in $B$ (``high–high''), whereas $z_A(i)<0$ and $z_{\mathrm{lag},B}(i)<0$ identify ``low–low'' regions that are suppressed with respect to both fields. Conversely, if $z_A(i)$ and $z_{\mathrm{lag},B}(i)$ have opposite signs, $I_i<0$ and the spaxel is locally ``inverted'' with respect to its surroundings: a positive $z_A(i)$ combined with a negative $z_{\mathrm{lag},B}(i)$ signals a high value of $A$ embedded in a depressed environment in $B$ (``high–low''), while the opposite configuration corresponds to a ``low–high'' structure. The magnitude $|I_i|$ quantifies the statistical significance of this local pattern in units of the product of the two scatters. The resulting $I_i$ map therefore provides a compact diagnostic of where, and how strongly, the field $A$ is locally correlated or anti-correlated with the spatial environment of the field $B$, enabling a spatially resolved assessment of concordant versus discordant structures in any pair of astrophysical quantities. In this study, because $A$ and $B$ are measured on the Voronoi tessellation and then mapped back onto the native spaxel grid, the resulting $I_i$ field is effectively smoothed on the bin scales but also suppresses structures and correlation signatures on scales smaller than a typical Voronoi bin size.

\section{Results}
\label{sec:results}




\subsection{The rMZR Relation}
\label{sec:rMZR_comparison}

\begin{figure*}
    \centering
    \includegraphics[width=\textwidth]{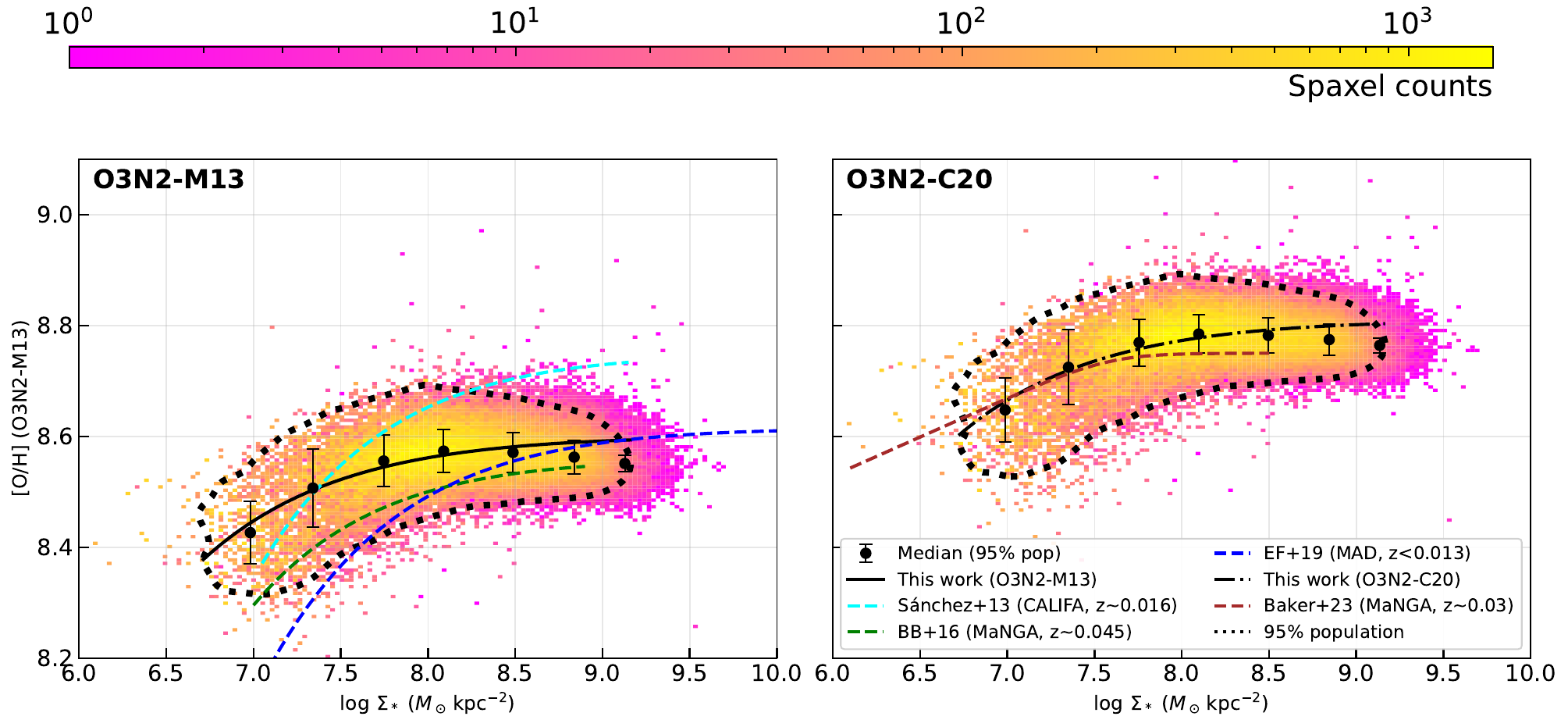}
    \caption{The resolved mass--metallicity relation (rMZR) for 13 MAUVE--MUSE galaxies (excluding NGC~4383). The left panel shows the rMZR derived using the \marino\ calibration, while the right panel shows the rMZR using the \curti\ calibration. The color scale indicates the number of spaxels. The dotted contours indicate the 95\% population bounds, and the black points with error bars show the median metallicity and standard deviation within the contours. The solid and dot-dashed black curves represent the best-fitting rMZR to our data using \autoref{eq:rMZR_fit} under \marino\ and \curti\ prescriptions, respectively. For comparison, we show literature relations: the blue dashed curve from MAD data \citep[$z<0.013$;][]{errozferrer19}, the cyan dashed curve from CALIFA data \citep[$z\sim0.016$;][]{sanchez13}, the green dashed curve from MaNGA data \citep[$z\sim0.045$;][]{barreraballesteros16} in the left panel, and the brown dashed curve from MaNGA data \citep[$z\sim0.03$;][]{baker23} in the right panel. Note that \citet{baker23} adopted the ``combined'' prescription from \citet{curti20}. }
    \label{fig:rMZR}
\end{figure*}

Before analysing the ensemble of resolved relations derived from combining all MAUVE--MUSE galaxies, we note that one of the 14 target galaxies, NGC~4383, is identified as a metallicity outlier relative to the rest of the sample. Across all four strong-line prescriptions adopted in this work (\dopita, \pg, \marino, and \curti), the oxygen abundances of star-forming spaxels in NGC~4383, especially the high-$\Sigma_\mathrm{SFR}$ spaxels (star-forming disc), are systematically offset low by at least 0.2 dex compared to the locus defined by the other 13 galaxies. This behaviour persists under our standard \textbf{HII} masking and across calibrators, and is therefore unlikely to be driven by methodological systematics. Because this single galaxy would otherwise dominate the low-metallicity tail and bias the ensemble trends, NGC~4383 is likely not a representative of the ``typical'' disk population in our sample. We therefore define a fiducial ensemble sample that excludes NGC~4383 for the purposes of fitting and benchmarking the ensemble trends. Hereafter, when we refer to `13 galaxies' or the `ensemble sample', NGC~4383 is excluded unless explicitly stated otherwise. We will return to discuss its properties in \autoref{sec:discussion_4383}. Additionally, the integrated properties of all 14 MAUVE--MUSE galaxies, including NGC~4383, are compared against the local universe's global scaling relations (SFMS, MZR, and FMR) in \aref{app:global_scaling}.

After combining all \textbf{HII}-selected spatial measurements from the 13 galaxies into a single ensemble, the resulting spaxel population follows the familiar rMZR shape reported in previous IFS surveys \citeg{sanchez13, barreraballesteros16, sanchez19, errozferrer19, yao22, baker23, koller24}: $Z_{\rm gas}$ increases steeply with $\Sigma_*$ at the low-mass end and gradually transitions to a shallower, saturation-like behaviour toward high $\Sigma_*$. In \autoref{fig:rMZR}, we present this locus using the 95\% population contour and its median trend (black points), shown for both the \marino\ (left) and \curti\ (right) calibrations. Within the 95\% contour, we perform a nonlinear least-squares fit by adopting the functional form from \citet{sanchez13} to describe the trend of rMZR:
\begin{equation}
    [\mathrm{O/H}] = a + b\,(\log_{10}\Sigma_* - c)\,e^{-(\log_{10}\Sigma_* - c)},
\label{eq:rMZR_fit}
\end{equation}
The best-fitting parameters for our 13 MAUVE--MUSE galaxies are listed in \autoref{tab:rMZR_fit_params}.

\begin{table}
\caption{Best-fitting parameters of the resolved MZR functional form in \autoref{eq:rMZR_fit} for the \marino\ and \curti\ metallicity prescriptions.}
\centering
\begin{tabular}{lccc}
\hline
Calibration & $a$ & $b$ & $c$ \\
\hline
\marino\ & $8.598 \pm 0.001$ & $0.00250 \pm 0.00018$ & $10.00 \pm 0.055$ \\
\curti\ & $8.807 \pm 0.001$ & $0.00237 \pm 0.00017$ & $10.00 \pm 0.056$ \\
\hline
\end{tabular}
\label{tab:rMZR_fit_params}
\end{table}

In the left panel of \autoref{fig:rMZR}, we compare our best-fitting rMZR obtained with the \marino\ calibration to three other spatially resolved local-universe surveys that adopt the same metallicity prescription. Overall, the MAUVE--MUSE relation shows the familiar saturating rMZR behaviour but exhibits the shallowest low-$\Sigma_*$ rise among the plotted studies. Relative to the MAD relation (dark blue dashed curve) of \citet{errozferrer19} at comparable ($\sim$100pc) resolution, we recover a similar high-$\Sigma_*$ plateau at $12+\log_{10}(\mathrm{O/H})\simeq 8.6$, whereas the MAD curve declines more steeply toward low $\Sigma_*$, reaching $\sim$0.3\,dex lower metallicity around $\log_{10}(\Sigma_*/M_\odot\,\mathrm{kpc}^{-2})\simeq 7.0$. Compared to the MaNGA-based relation (green dashed curve) of \citet{barreraballesteros16}, our rMZR is systematically higher by $\sim$0.05--0.1\,dex over the entire dynamic range of $\log_{10}(\Sigma_*)$. The comparison to CALIFA relation (cyan dashed curve) by \citet{sanchez13} is non-monotonic: in the intermediate- and high-mass regimes ($7.5\lesssim \log_{10}(\Sigma_*/M_\odot\,\mathrm{kpc}^{-2}) \lesssim 9.0$) our curve lies below \citet{sanchez13}, but at $\log_{10}(\Sigma_*/M_\odot\,\mathrm{kpc}^{-2})\lesssim 7.5$ it becomes higher, so that our rMZR is consistent with \citet{barreraballesteros16} and implies a comparatively shallow low-$\Sigma_*$ metallicity decline in MAUVE--MUSE data. Part of this offset may reflect differences in sample selection: the early CALIFA analysis of \citet{sanchez13} is effectively complete only above $\log_{10}(M_*/M_\odot)\gtrsim9.5$. As a result, its high-$\Sigma_*$ component is weighted toward more massive, more metal-rich discs, which could elevate the CALIFA normalization relative to our late-type Virgo spirals in MAUVE.

The right panel shows that adopting \curti\ shifts the best-fitting rMZR upward by $\sim$0.2\,dex relative to \marino\ at fixed $\Sigma_*$, which is consistent with the findings from \citet{scudder21}, underscoring that the absolute normalisation is calibration-dependent (see further discussion in \autoref{sec:discussion_metallicity}). In this representation, the ``combined'' prescription used by \citet{baker23} yields a best-fitting relation (brown dashed curve) that is highly consistent with our \curti\ fit, particularly at $\log_{10}(\Sigma_*/M_\odot\,\mathrm{kpc}^{-2})\lesssim 7.5$. Aditionally, although not shown here, our \marino\ relation has a similar overall curvature to other IFS measurements at larger distances and/or intermediate redshifts \citep[e.g.,][]{yao22,koller24} but with a higher normalisation; as discussed by \citet{koller24}, such offsets are unlikely to be explained by redshift evolution of the rMZR alone. Finally, as noted by \citet{koller24}, we acknowledge that differences in $\Sigma_*$ estimation (i.e. SSP assumptions) and related pipeline choices can also contribute to survey-to-survey rMZR discrepancies.

\subsection{rMZR's Secondary Dependence of rMZR on $\Sigma_\mathrm{SFR}$}

\begin{figure*}
    \centering
    \includegraphics[width=\textwidth]{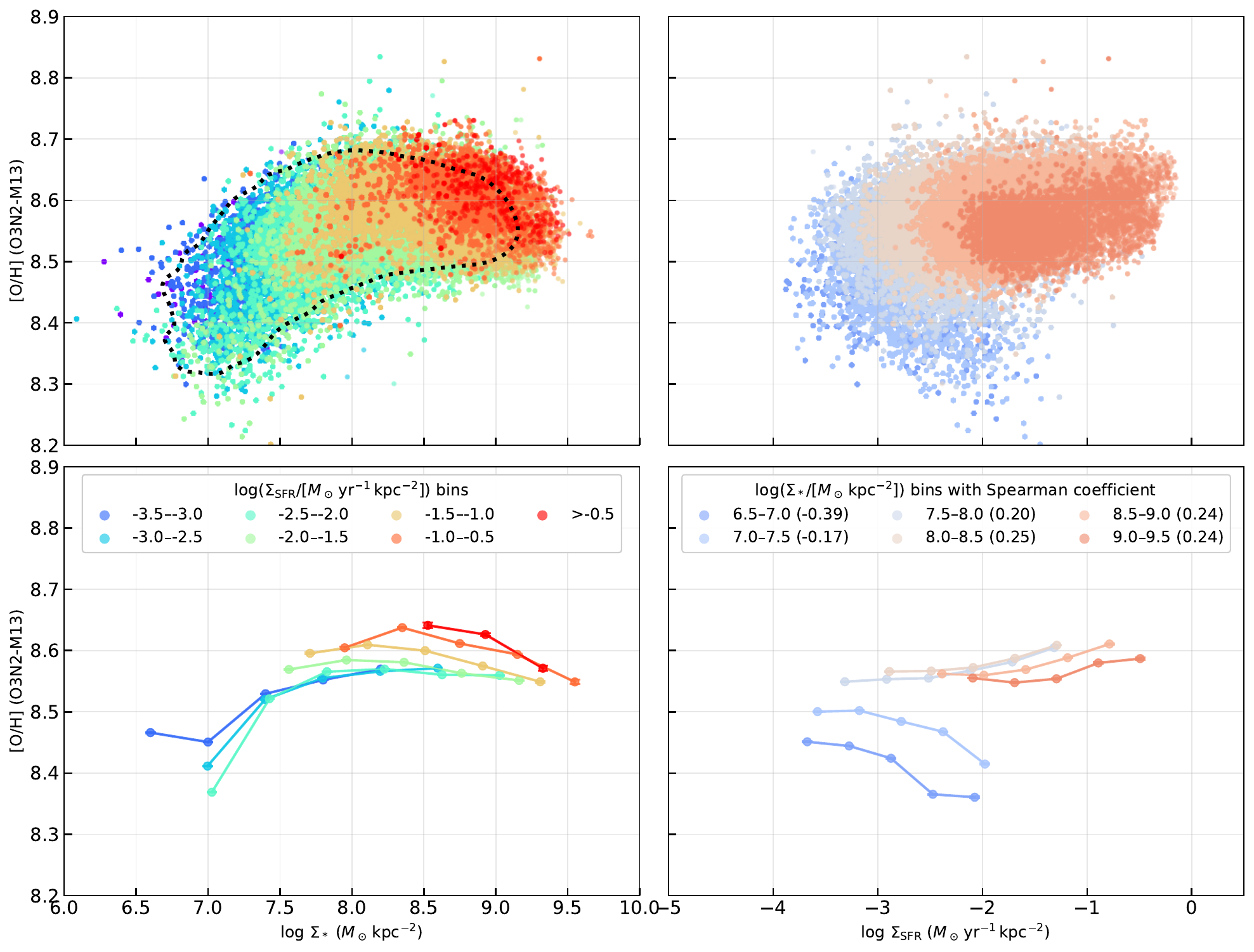}
    \caption{The left column shows the rMZR colour-coded by $\Sigma_\mathrm{SFR}$ for 13 MAUVE--MUSE galaxies (excluding NGC~4383), while the right column shows the $Z_\mathrm{gas}$--$\Sigma_\mathrm{SFR}$ relation colour-coded by $\Sigma_*$. The upper panels display the scatter plots, while the lower panels show the corresponding median trends of each scatter plot. Similar to \autoref{fig:rMZR}, we use the black contour to enclose the 95\% population of \textbf{HII} spaxels. Within this contour, we present the median trends with errorbars indicating the standard error of the median (SEM; typical value of $\lesssim 0.01$).  Note that due to insufficient statistics, the median trends of the lowest and highest $\Sigma_*$ bins in the bottom right panel are not shown.}
    \label{fig:main}
\end{figure*}

Colour-coding the rMZR by $\Sigma_\mathrm{SFR}$ reveals a clear secondary dependence on local star formation intensity (see the left column of \autoref{fig:main}). At fixed $\Sigma_*$, high-$\Sigma_\mathrm{SFR}$ spaxels tend to lie at higher $Z_\mathrm{gas}$ in the high-$\Sigma_*$ regime, whereas the ordering reverses at low $\Sigma_*$, where lower-$\Sigma_\mathrm{SFR}$ spaxels are preferentially more metal-rich. The transition between these regimes occurs at $\log_{10}(\Sigma_*/M_\odot\,\mathrm{kpc}^{-2}) \simeq 7.5$--$8.0$. While the exact reversal point is imprecise due to limited statistics in the narrow overlap region, its location is consistent with the inversion reported by MAGPI by \citet{koller24}, who found a clear sign change near $\log_{10}(\Sigma_*/M_\odot\,\mathrm{kpc}^{-2})\sim 7.8$ using \marino\ prescription as well. The MAUVE--MUSE data therefore support a mass-dependent inversion in the resolved $Z_\mathrm{gas}$--$\Sigma_\mathrm{SFR}$ coupling, suggesting that the dominant driver of the secondary dependence (the balance between metal enrichment and dilution) may differ between low- and high-$\Sigma_*$ environments. Similarly, although not a strong signal, under the prescriptions of \citet{curti20}, MaNGA-based analysis of \citet{baker23} also appears consistent with a flipped correlation in the spatially resolved scenario, with the transition occurring at $\log_{10}(\Sigma_*/M_\odot\,\mathrm{kpc}^{-2})\approx 7.5$--$7.7$ (their Figure 2), in good agreement with our inferred range. Furthermore, adopting both \dopita\ and \pg\ calibrations, \wl\ reported a predominantly positive local $Z_\mathrm{gas}$--$\Sigma_\mathrm{SFR}$ correlation using the MAD data at $\sim 100$\,pc resolution, whereas with MaNGA data on global galactic scales they found an overall negative correlation. 

To visualise metallicity's dependence on $\Sigma_\mathrm{SFR}$ more directly, the right column of \autoref{fig:main} shows $Z_\mathrm{gas}$ as a function of $\Sigma_\mathrm{SFR}$ in bins of $\Sigma_*$. For each $\Sigma_*$ bin, we also compute the Spearman rank correlation coefficient $\rho$ between $Z_\mathrm{gas}$ and $\Sigma_\mathrm{SFR}$ after removing the $3\sigma$ outliers. The lowest-$\Sigma_*$ bins ($\log_{10}\Sigma_*$ at 6.5 - 7.0) exhibit the strongest anti-correlation at Spearman coefficient $\rho=-0.41$, with metallicity decreasing as $\Sigma_\mathrm{SFR}$ rises. With increasing $\Sigma_*$, the median trends flatten progressively, and ultimately the relation becomes positively correlated, such that regions of elevated star formation are also metal-enhanced. This systematic, mass-ordered sequence from negative through flat to positive slopes encapsulates the rMZR's second dependence on $\Sigma_\mathrm{SFR}$ and motivates the offset and spatial-correlation analyses undertaken in the next section.

\subsection{Offset Relation and Local Moran-like Correlation Map}

\begin{figure*}
    \centering
    \includegraphics[width=\textwidth]{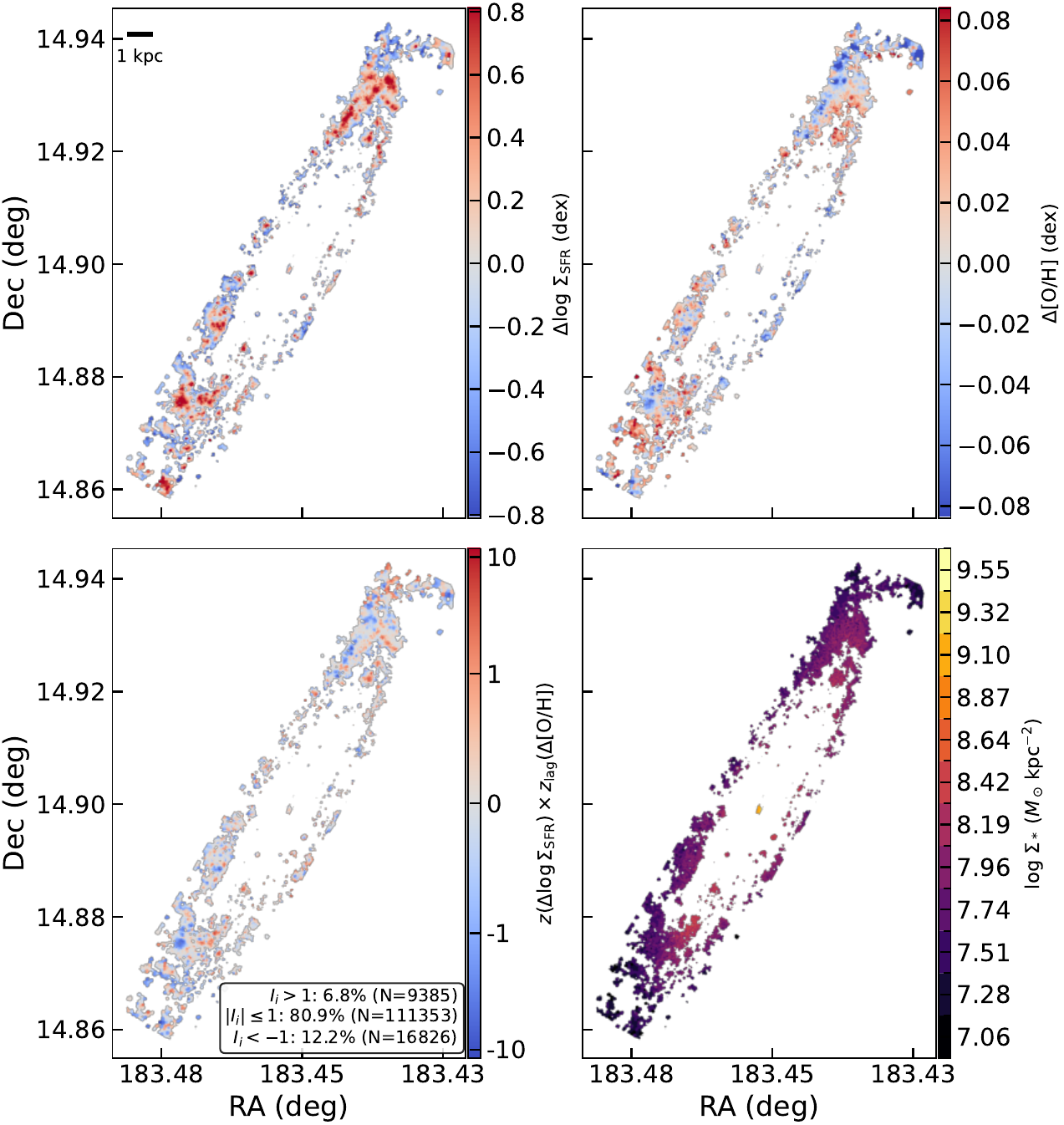}
    \caption{Example offset and local Moran-like correlation maps for \textbf{HII} spaxels of NGC~4192. The top panels show the offsets $\Delta\log_{10}\Sigma_{\rm SFR}$ (left) and $\Delta[{\rm O/H}]_{\rm O3N2-M13}$ (right) after removing the mean trends with $\Sigma_*$ (see text). The bottom-left panel presents the local bivariate Moran-like product $I_i$ constructed from these two offset fields, highlighting spatially local correlation (red) and anti-correlation (blue). The bottom-right panel shows $\log_{10}\Sigma_*$ for the \textbf{HII} spaxels used in the other panels, divided into 12 equal-width bins.}
    \label{fig:4192}
\end{figure*}

\begin{figure*}
    \centering
    \includegraphics[width=\textwidth]{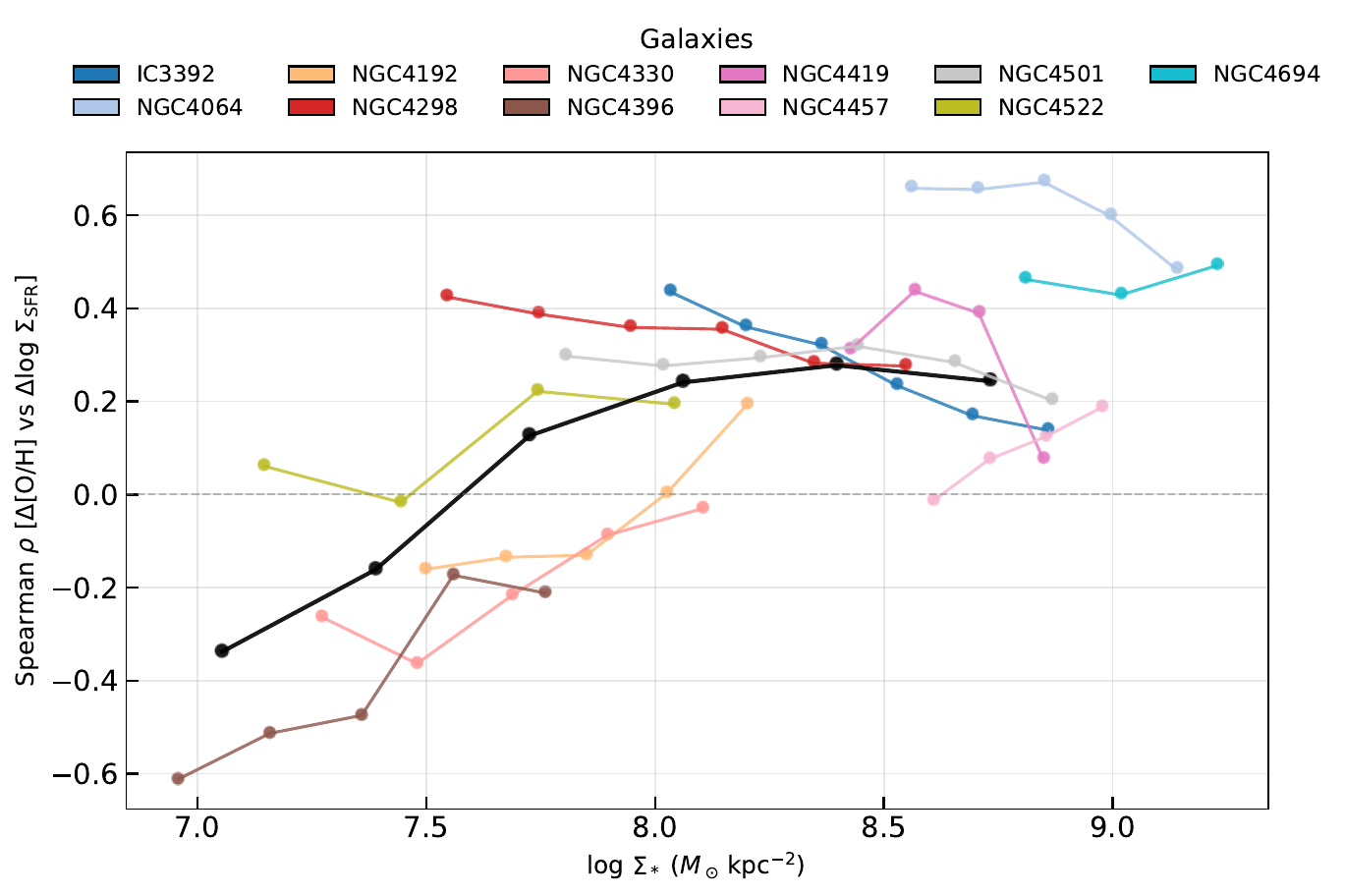}
    \caption{Spearman rank correlation coefficient $\rho$ between $\Delta[{\rm O/H}]_{\rm O3N2-M13}$ and $\Delta\log_{10}\Sigma_{\rm SFR}$ as a function of $\log_{10}\Sigma_*$, measured in discrete $\Sigma_*$ bins for 13 MAUVE--MUSE galaxies (excluding NGC~4383). Each coloured curve corresponds to one galaxy; the Spearman trends of NGC~4293 and NGC~4698 are not shown here due to insufficient statistics. The black curve shows the ensemble result obtained by combining all \textbf{HII} spaxels from the 13 galaxies.}
    \label{fig:spearman}
\end{figure*}

To isolate the secondary coupling between metallicity and star formation from their primary dependence on stellar mass surface density, we construct mass-normalised offset maps for each galaxy. Specifically, within each galaxy we divide all valid \textbf{HII} spaxels into 12 equal-width bins in $\log_{10}\Sigma_*$ (illustrated for one of MAUVE--MUSE targets, NGC~4192, in the bottom-right panel of \autoref{fig:4192}). For every bin $m$, we compute the mean spaxel values $\langle \log_{10}\Sigma_{\rm SFR}\rangle_m$ and $\langle [{\rm O/H}]_{\rm O3N2-M13}\rangle_m$, and define spaxel-level offsets
\begin{equation}
    \begin{cases}
        \Delta\log_{10}\Sigma_{\rm SFR}(i) \equiv \log_{10}\Sigma_{\rm SFR}(i) - \langle \log_{10}\Sigma_{\rm SFR}\rangle_m, \\
        \Delta[{\rm O/H}](i) \equiv [{\rm O/H}](i) - \langle [{\rm O/H}]\rangle_m,
    \end{cases}
    \label{eq:offset}
\end{equation}
where spaxel $i$ belongs to bin $m$. By construction, these offsets remove the mean rMZR and resolve star forming main sequence (rSFMS) trends within each $\Sigma_*$ interval, leaving only the relative fluctuations of $\Sigma_{\rm SFR}$ and $Z_{\rm gas}$ at fixed local stellar mass density. We note that residual correlations can in principle be induced by measurement uncertainties and covariances when subtracting primary scaling relations \citep[see Section 3.4 of][]{sanchez21b}; however, in our analysis the offsets are defined and the correlations are measured independently within narrow $\Sigma_*$ bins. The resulting maps therefore trace spatial variations that are not driven by the underlying $\Sigma_*$ distribution and its associated large-scale gradients (e.g., see top panels of \autoref{fig:4192}).

Using these mass-normalised quantities, we also measure the residual $Z_{\rm gas}$--$\Sigma_{\rm SFR}$ coupling by computing, in each $\Sigma_*$ bin of each galaxy, the Spearman rank correlation coefficient between $\Delta[{\rm O/H}]_{\rm O3N2-M13}$ and $\Delta\log_{10}\Sigma_{\rm SFR}$ after removing the $3\sigma$ outliers. The resulting $\rho(\Sigma_*)$ trends are shown in \autoref{fig:spearman}; each reported $\rho$ has a corresponding p-value $\lesssim 0.01$, implying a statistically significant correlation between the two quantities. For the ensemble sample (black curve), $\rho$ increases smoothly from negative values at low $\Sigma_*$ to positive values at high $\Sigma_*$: rising from $\rho\!\approx\!-0.35$ at $\log_{10}\Sigma_*\!\sim\!7$ to $\rho\!\approx\!+0.3$ at $\log_{10}\Sigma_*\!\gtrsim\!8$, with a zero-crossing within $\log_{10}(\Sigma_*/M_\odot\,{\rm kpc}^{-2})\simeq 7.5$--$8.0$. Individual galaxies follow the same qualitative ordering: systems that predominantly sample the low-$\Sigma_*$ regime (e.g.\ NGC~4396, NGC~4457, and part of NGC~4192) contribute to the anti-correlated branch, whereas galaxies whose \textbf{HII} spaxels lie mainly at high $\Sigma_*$ populate the positively correlated regime. Notably, NGC~4192 spans a broad $\Sigma_*$ range that straddles the inversion point and exhibits a clear internal sign reversal, indicating that the transition is intrinsic to the subtle physics within galaxies rather than an artefact of combining different galaxy populations.

We further assess the spatial locality of this residual coupling by constructing the local bivariate Moran-like product $I_i$ defined in \autoref{sec:moran}. As an illustrative example, the bottom-left panel of \autoref{fig:4192} shows the $I_i$ map for NGC~4192, derived from the corresponding $\Delta\log_{10}\Sigma_{\rm SFR}$ and $\Delta[{\rm O/H}]$ fields. As indicated by \autoref{eq:offset}, these two fields are defined as residuals at fixed $\Sigma_*$, i.e.\ after subtracting, in each $\Sigma_*$ bin, the trends of mean rMZR and rSFMS. In practice, this removes the dominant radial (mass-driven) profiles, so that the $I_i$ map traces purely local spatial correspondence between star formation and metallicity fluctuations. By construction, $I_i$ is the product of two standardised quantities, $z_{\Delta\log_{10}\Sigma_{\rm SFR}}(i)$ and $z_{\mathrm{lag},\Delta[{\rm O/H}]}(i)$, each expressed in units of each $\Sigma_*$ bin's standard deviation of the corresponding field. Thus $|I_i|\sim 1$ corresponds to the joint product of two $\sim 1\sigma$ deviations (one at the spaxel position, one in its local environment), whereas $|I_i|\ll 1$ indicates that at least one of the two is close to the global mean. Therefore, while the choice is inherently arbitrary, we adopt $|I_i|=1$ as a convenient, dimensionless threshold to distinguish weak or typical patterns from stronger local correspondence. 

For NGC~4192, most spaxels in \autoref{fig:4192} have $|I_i|\le 1$ (grey), implying weak or ambiguous local correspondence, while a non-negligible fraction ($\sim20\%$) show significant local correlation ($I_i>1$, red: higher (lower) $\Sigma_\mathrm{SFR}$ with higher (lower) $Z_{\rm gas}$) or anti-correlation ($I_i<-1$, blue: higher (lower) $\Sigma_\mathrm{SFR}$ with lower (higher) $Z_{\rm gas}$). These two tails are of comparable magnitude, with $6.8\%$ of spaxels having $I_i>1$ and $12.2\%$ having $I_i<-1$, while the majority ($80.9\%$) remain in the weak/ambiguous regime. This near-balance between correlated and anti-correlated \ion{H}{ii} spaxels is consistent with NGC~4192 being the clearest ``inversion'' case in our sample: depending on $\Sigma_*$, the net $Z_{\rm gas}$--$\Sigma_{\rm SFR}$ coupling can change sign in \autoref{fig:spearman}. Interpreting high-$\Sigma_{\rm SFR}$ peaks as tracing the cores of individual \ion{H}{ii} regions, the coexistence of red and blue structures within the same $\Sigma_*$ bins suggests that the local $Z_{\rm gas}$--$\Sigma_{\rm SFR}$ coupling is spatially heterogeneous: once mass trends of these two properties are removed, neighbouring regions of comparable $\Sigma_*$ can display either concordant or discordant metallicity responses to star formation. Moreover, the highest $|I_i|$ values are spatially coincident with bright H$\alpha$/high-$\Sigma_{\rm SFR}$ clumps that satisfy our \textbf{HII} \bpt\ selection, whereas diffuse low-$\Sigma_{\rm SFR}$ regions tend to show $|I_i|\lesssim 1$, supporting that the observed signal is associated with genuine \ion{H}{ii} regions (or clumps of them) rather than being driven by DIG contamination. These maps show that (i) the sign of the $\Delta Z_{\rm gas}$--$\Delta\Sigma_{\rm SFR}$ coupling is set by heterogeneous local physical processes, so correlated and anti-correlated \ion{H}{ii} regions can coexist at fixed $\Sigma_*$, even in neighboring regions; and (ii) the observed mass-dependent behaviour arises only as an emergent population trend when one aggregates enough spaxels to sample both sides of the transition around $\log_{10}(\Sigma_*/M_\odot\,{\rm kpc}^{-2})\sim 7.5$–8.0.  

\section{The Origin of $Z_\mathrm{gas}$--$\Sigma_\mathrm{SFR}$ (Anti-)Correlation in a Gas-regulator Framework}
\label{sec:origin}

To interpret the varying sign of the $Z_\mathrm{gas}$--$\Sigma_\mathrm{SFR}$ relation and its residual $\Delta Z_{\rm gas}$--$\Delta\Sigma_{\rm SFR}$ coupling, we adopt a spatially resolved gas–regulator (``bathtub'') framework formulated entirely in spaxel-by-spaxel variations. Throughout this section, all quantities are written as surface densities to reflect the fact that our observational constraints are local ($\sim 100$\,pc resolution) rather than global. The model is intended to describe the evolution of the gas regulator in the local universe, over timescales that are short compared to the characteristic orbital period in the disc ($\sim200 - 300$ Myr). In practice, the variations that matter for our observed $Z_\mathrm{gas}$--$\Sigma_\mathrm{SFR}$ (anti-)correlation are closer to the lifetimes of individual \ion{H}{ii} regions and local dynamical or mixing timescales (i.e.\ a few tens of Myr), as suggested by the fact that the strongest signal (strongest correlation between $\mathrm{[O/H]}$ and $\Sigma_\mathrm{SFR}$) is associated with bright H$\alpha$/high-$\Sigma_\mathrm{SFR}$ clumps. Over such intervals, the accumulated growth of the stellar mass surface density $\Sigma_*$ is negligible compared to its present-day value (or at most comparable to observational uncertainties $\lesssim0.05$ dex), so we neglect the time evolution of $\Sigma_*$ and treat it as a fixed control parameter. We therefore do not attempt to reconstruct the full past star formation or metal enrichment history of each region. Instead, we assume that within each galaxy, and at fixed $\Sigma_*$, different \ion{H}{ii} spaxels within a galaxy began their recent evolution from broadly similar initial conditions and have since evolved under the same set of governing physical conditions. 

Within this framework, the time variable, $t$, plays the role of an evolution parameter that controls the instantaneous state of the local regulator (gas mass, SFR, metallicity, gas supply properties) at a given location. Mathematically, even though we observe different spaxels at the same cosmic time, each with its own set of properties, $P$ (here $P\in\{\Sigma_*,\Sigma_\mathrm{SFR}, Z_\mathrm{gas},\mathrm{etc.}\}$), it is useful to think of these spatial variations as sampling different points along a common family of time-dependent solutions. Equivalently, given any pair of spaxels within the same $\Sigma_*$ bin, the differences in their local properties can be interpreted as the result of having evolved for different effective ``phases'' or having experienced different recent perturbations under the same functional form of the regulator equations. In this sense, the time coordinate, $t$, may be regarded as a parameter that indexes the instantaneous value of each property and allows us to trace how fluctuations in one property, $P$, induce corresponding responses in another. What follows is therefore a local, time-dependent gas-regulator model for the coupled variations of $\Sigma_{\rm gas}$, $\Sigma_{\rm SFR}$, and $Z_{\rm gas}$ at fixed $\Sigma_*$, which we use as a physical language for organizing the spatially resolved fluctuations measured in the MAUVE--MUSE data.

We express the star-formation rate surface density as
\begin{equation}
    \Sigma_\mathrm{SFR}(t) \equiv \frac{\Sigma_\mathrm{g}(t)}{\tau_\mathrm{dep}(t)},
\label{eq:Sigma_SFR}    
\end{equation}
where $\Sigma_\mathrm{g}(t)$ is the molecular gas surface density and $\tau_\mathrm{dep}(t)$ the local depletion time. Then, we use the following two differential equations to describe our spatially resolved framework. First, the gas mass surface density obey
\begin{equation}
    \frac{d\Sigma_\mathrm{g}(t)}{dt}
    = \Sigma_\Phi(t) - (1-R)\Sigma_\mathrm{SFR}(t)-\Sigma_\mathrm{out}(t),
\end{equation}
where $\Sigma_\Phi(t)$ is the \emph{gas supply rate (GSR) surface density}\footnote{We avoid the term ``inflow rate surface density'' because inflow usually denotes pristine, galaxy-scale accretion from the circumgalactic medium (CGM), whereas on local scales it is more appropriate to describe a region's or giant molecular cloud (GMC)'s ability to gain gas supply from its ambient ISM; moreover, for our observed Virgo members, the global inflow is likely suppressed or truncated \citep[starvation;][]{larson80}.}, i.e.\ the rate at which the region gains fresh gas per unit area, $R$ is the fraction of the stellar mass which is immediately returned by massive stars, and $\Sigma_\mathrm{out}(t)$ is the outflow rate surface density which accounts for gas removal driven by both internal feedbacks and environmental mechanisms like stripping. 

Then, the surface density of metals ($\Sigma_Z$) follows
\begin{equation}
\begin{split}
    \frac{d\Sigma_Z(t)}{dt}
    &= y(1-R)\,\Sigma_\mathrm{SFR}(t)
      - \bigl[(1-R)\Sigma_\mathrm{SFR}(t)+\Sigma_\mathrm{out}(t)\bigr]Z(t) \\
    &\quad + \Sigma_\Phi(t)\,Z_\Phi,
\end{split}
\end{equation}
with metal yield per stellar mass, $y$, and metallicity of the supplied gas, $Z_\Phi$ (i.e. background metallicity). 

Defining the gas-phase metallicity as
\begin{equation}
    Z(t)\equiv \frac{\Sigma_Z(t)}{\Sigma_\mathrm{g}(t)},
\end{equation}
we obtain
\begin{equation}
    \frac{dZ(t)}{dt}
    = \frac{1}{\Sigma_\mathrm{g}(t)}\left[
        y(1-R)\,\Sigma_\mathrm{SFR}(t)
        + \bigl(Z_\Phi - Z(t)\bigr)\,\Sigma_\Phi(t)
      \right],
\label{eq:dZdt_general_origin}
\end{equation}
so that at any location, the metallicity responds to enrichment from ongoing star formation and dilution by the supplied gas.

It is convenient to introduce the gas supply timescale
\begin{equation}
    \tau_\Phi(t) \equiv \frac{\Sigma_\mathrm{g}(t)}{\Sigma_\Phi(t)}
    \ \le\ \tau_\mathrm{dep}(t),
\label{eq:tau_phi}
\end{equation}
which measures how quickly the local gas reservoir would be replenished by the current GSR. Using \autoref{eq:Sigma_SFR} and \autoref{eq:tau_phi}, \autoref{eq:dZdt_general_origin} yields
\begin{equation}
    \frac{dZ(t)}{dt}
    = \frac{1}{\tau_\Phi(t)}
      \left[
        Z_\Phi + y(1-R)\frac{\Sigma_\mathrm{SFR}(t)}{\Sigma_\Phi(t)} - Z(t)
      \right].
\label{eq:dZdt_driver_form}
\end{equation}
This motivates the definition of the \emph{instantaneous target metallicity}:
\begin{equation}
    Z^\dagger(t) \equiv Z_\Phi + y(1-R)\frac{\Sigma_\mathrm{SFR}(t)}{\Sigma_\Phi(t)},
\label{eq:Zdagger_definition}
\end{equation}
so that \autoref{eq:dZdt_driver_form} becomes
\begin{equation}
    \frac{dZ(t)}{dt}
    = \frac{1}{\tau_\Phi(t)}\bigl[Z^\dagger(t) - Z(t)\bigr]
    \quad\Longleftrightarrow\quad
    Z^\dagger(t) = Z(t) + \frac{dZ(t)}{dt}\,\tau_\Phi(t).
\label{eq:Zdagger_driver_relation}
\end{equation}
This explicitly shows that $Z^\dagger(t)$ acts as the \emph{driver} of the metallicity evolution compared to the real-time metallicity $Z(t)$ (the observed metallicity): $Z^\dagger(t)$ is the metallicity that the system would reach if the current rate of change $dZ/dt$ were sustained over one gas supply timescale. Only when $Z^\dagger(t)=Z(t)$ does \autoref{eq:Zdagger_driver_relation} imply $dZ/dt=0$, corresponding to a \emph{local minimum or maximum} of $Z(t)$ rather than a long-lived equilibrium.

With these definitions, the observed spaxels can be separated conceptually into two complementary cases: either reaching (\autoref{sec:case1}) or evolving toward/away from (\autoref{sec:case2}) the local extremum stages. \autoref{tab:regulator_symbols} summarises the notation used in this spatially resolved gas-regulator model. 

\begin{table}
    \caption{Symbols used in the spatially resolved gas-regulator model.}
    \centering
    \begin{tabular}{l p{0.72\columnwidth}}
    \hline
    Symbol & Meaning \\
    \hline
    \hline
    $t$ & Time / evolution parameter indexing the instantaneous state of the local regulator (used to organise spaxel-to-spaxel fluctuations).\\
    \hline
    $\Sigma_*$ & Stellar mass surface density; treated as a fixed control parameter within a given $\Sigma_*$ bin over the short timescales of interest.\\
    \hline
    $\Sigma_{\rm g}(t)$ & Gas mass surface density (the local molecular gas reservoir).\\
    \hline
    $\Sigma_{\rm SFR}(t)$ & Star formation rate (SFR) surface density.\\
    \hline
    $\tau_{\rm dep}(t)$ & Gas depletion time, defined by $\Sigma_{\rm SFR}\equiv \Sigma_{\rm g}/\tau_{\rm dep}$.\\
    \hline
    $\Sigma_{\Phi}(t)$ & Gas supply rate (GSR) surface density: gas accretion rate at which the region gains gas per unit area from its surroundings (local supply/replenishment).\\
    \hline
    $\tau_{\Phi}(t)$ & Gas supply timescale, $\tau_{\Phi}\equiv \Sigma_{\rm g}/\Sigma_{\Phi}$ (replenishment time of the local gas reservoir at the current GSR).\\
    \hline
    $\Sigma_{\rm out}(t)$ & Outflow rate surface density (gas removed by feedback and/or environmental processes such as stripping).\\
    \hline
    $R$ & Return fraction: fraction of newly formed stellar mass promptly recycled back to the ISM.\\
    \hline
    $\Sigma_{Z}(t)$ & Metal surface density in the gas phase (in the same metal species traced by $Z$; here effectively oxygen).\\
    \hline
    $Z(t)$ & Gas-phase metallicity of the local reservoir, $Z\equiv \Sigma_Z/\Sigma_{\rm g}$ (model variable corresponding to an oxygen abundance).\\
    \hline
    $y$ & Metal yield per unit stellar mass formed.\\
    \hline
    $Z_{\Phi}$ & Metallicity of the supplied gas (background/pre-enrichment level of the gas entering the region).\\
    \hline
    $Z^{\dagger}(t)$ & Instantaneous target metallicity (driver/attractor for $Z$), $Z^{\dagger}\equiv Z_{\Phi}+y(1-R)\,\Sigma_{\rm SFR}/\Sigma_{\Phi}$.\\
    \hline
    \end{tabular}
    \label{tab:regulator_symbols}
\end{table}

\subsection{Case I: Correlation at $Z=Z^\dagger$}
\label{sec:case1}

For spaxels observed at a series of local extremum times $t=t_m$, the real-time metallicity satisfies $dZ/dt|_{t_m}=0$, hence $Z(t_m)=Z^\dagger(t_m)$, with
\begin{equation}
    Z^\dagger(t_m)
    = Z_\Phi + y(1-R)\frac{\Sigma_\mathrm{SFR}(t_m)}{\Sigma_\Phi(t_m)} .
\label{eq:Zdagger_tm_simple}
\end{equation}
Taking a first-order expansion of \autoref{eq:Zdagger_tm_simple} yields
\begin{equation}
    \delta Z^\dagger(t_m)
    = \bigl[Z^\dagger(t_m)-Z_\Phi\bigr]
      \left[
        \delta\log_{10}\Sigma_\mathrm{SFR}(t_m)
        - \delta\log_{10}\Sigma_\Phi(t_m)
      \right],
\label{eq:deltaZdagger_tm}
\end{equation}
so the sign of the metallicity fluctuation at extrema is controlled by the competition between spatial variations in SFR and GSR. If $\Sigma_\mathrm{SFR}$ varies more strongly than $\Sigma_\Phi$ across neighbouring spaxels, then $\delta Z^\dagger$ follows $\delta\log_{10}\Sigma_\mathrm{SFR}$ and a positive $Z_\mathrm{gas}$--$\Sigma_\mathrm{SFR}$ correlation is expected. If instead GSR variations dominate while $\Sigma_\mathrm{SFR}$ and $\Sigma_\Phi$ co-vary, dilution drives $\delta Z^\dagger$ opposite to $\delta\log_{10}\Sigma_\mathrm{SFR}$, yielding an anti-correlation. Where the two contributions are comparable, $\delta Z^\dagger$ is small and little correlation is produced.

Using \autoref{eq:Sigma_SFR} and \autoref{eq:tau_phi}, \autoref{eq:deltaZdagger_tm} is equivalently
\begin{equation}
    \delta Z^\dagger(t_m)
    \ \propto\
    \delta\log_{10}\tau_\Phi(t_m)
    - \delta\log_{10}\tau_\mathrm{dep}(t_m),
\label{eq:deltaZdagger_tm_tau}
\end{equation}
highlighting that, at metallicity extrema, the correlation sign is set by whether variations in the gas supply timescale outpace those in depletion time. Furthermore, if we generalize (i.e. substitute) $t_m$ with $t$, which in fact describes the change of instantaneous target metallicity rather than the real-time metallicity at extrema, we still have the same functional forms of \autoref{eq:deltaZdagger_tm} and \autoref{eq:deltaZdagger_tm_tau}. This tells us that the change of instantaneous target metallicity is also determined by the comparison of the relative change magnitude between the local SFR and GSR. We will return to this relation later in our discussion. 

\subsection{Case II: Correlation at $Z\neq Z^\dagger$}
\label{sec:case2}

In reality, most spaxels are not observed exactly at a local metallicity extremum. To relate their real-time metallicity to the driver $Z^\dagger$, we expand around the nearest extremum time $t_m$ by rewriting $t=t_m+\zeta$ ($\zeta$ is an arbitrary time interval away from the extremum time $t_m$) and, for any quantity $P$ (i.e. $P\in\{\Sigma_\mathrm{SFR},Z,Z^\dagger,\mathrm{etc.}\}$), defining
\begin{equation}
    \Delta_P(\zeta)\equiv P(t_m+\zeta)-P(t_m).
\end{equation}
Linearising \autoref{eq:Zdagger_driver_relation} about $t_m$ yields a first-order response,
\begin{equation}
    \frac{d\,\Delta_Z(\zeta)}{d\zeta}
    + \frac{\Delta_Z(\zeta)}{\tau_\Phi(t_m)}
    = \frac{\Delta_{Z^\dagger}(\zeta)}{\tau_\Phi(t_m)} ,
\end{equation}
whose solution is
\begin{equation}
    \Delta_Z(\zeta)
    = \frac{1}{\tau_\Phi(t_m)}
      \int_0^\zeta
      \exp\!\left[-\frac{\zeta-\nu}{\tau_\Phi(t_m)}\right]
      \Delta_{Z^\dagger}(\nu)\,d\nu .
\label{eq:convolution_case2}
\end{equation}
\autoref{eq:convolution_case2} shows that $\Delta_Z(\zeta)$ is a low-pass-filtered version of the driver $\Delta_{Z^\dagger}(\zeta)$: the real-time metallicity responds as an exponential moving average of the target metallicity, smoothed over the gas supply timescale $\tau_\Phi(t_m)$. Physically, this ``memory'' arises because $Z(t)$ cannot respond instantaneously to changes in $Z^\dagger(t)$, due to finite delays in metal injection from recent star formation and in turbulent mixing (metal diffusion) through the ISM. In practice, because of detection procedures and measurement uncertainties, our observations predominantly probe the regime $\zeta \gtrsim \tau_\Phi(t_m)$, i.e.\ the time-averaged response rather than an instantaneous snapshot of $Z^\dagger$.

For such observed spaxels, $\delta Z(t)$ follows the recent behaviour of $\delta Z^\dagger(t)$ and therefore inherits the same sign set by star formation versus gas supply variability. Inserting \autoref{eq:convolution_case2} into \autoref{eq:Zdagger_driver_relation}, yields the consistent prediction as \autoref{eq:deltaZdagger_tm} and \autoref{eq:deltaZdagger_tm_tau}: 
\begin{equation}
\begin{split}
    \delta Z(t)
    &\propto
    \delta\log_{10}\!\bigl(\Sigma_\mathrm{SFR}(t)\bigr)
    - \delta\log_{10}\!\bigl(\Sigma_\Phi(t)\bigr) \\
    &\propto
    \delta\log_{10}\!\bigl(\tau_\Phi(t)\bigr)
    - \delta\log_{10}\!\bigl(\tau_\mathrm{dep}(t)\bigr).
\end{split}
\label{eq:deltaZ_case2_final}
\end{equation}
Thus, away from extrema, the observed local $Z_\mathrm{gas}$--$\Sigma_\mathrm{SFR}$ (anti-)correlation is still governed by whether spatial variations are driven primarily by changes in star-formation efficiency (through $\tau_\mathrm{dep}$) or by changes in the gas supply rate (through $\tau_\Phi$), with the finite response time encoded in $\tau_\Phi$ introducing additional scatter and spatially varying phase lags.

\subsection{Interpretation of the Observed (Anti-)Correlation}

Within this framework, our resolved MAUVE--MUSE trends in \autoref{fig:spearman} can be interpreted as the outcome of a competition between GSR-driven and SFR-driven variability of each gas regulator. Observationally, in high-$\Sigma_*$ regions (typically inner discs), dense gas and short depletion times make spatial variations in $\Sigma_\mathrm{SFR}$ or $\tau_\mathrm{dep}$ relatively strong, while the ability to gain fresh gas (i.e. the GSR) is comparatively steady from spaxel to spaxel. In this regime, \autoref{eq:deltaZ_case2_final} favours a positive $Z_\mathrm{gas}$--$\Sigma_\mathrm{SFR}$ correlation: fluctuations in SFR dominate over variations in GSR, and oxygen abundance increases with $\Sigma_\mathrm{SFR}$ at fixed $\Sigma_*$. On the other hand, at low $\Sigma_*$ (outer discs), especially in the Virgo environment where the local gas supply can fluctuate strongly due to environmental processes, spatial variations in $\tau_\Phi$ become more important. When GSR fluctuations dominate while $\Sigma_\mathrm{SFR}$ and $\Sigma_\Phi$ co-vary, the dilution associated with enhanced supply changes $Z_\mathrm{gas}$ to lower values in regions of relatively higher SFR, driving the correlation toward zero or negative values. In between these extremes, where variations in $\tau_\Phi$ and $\tau_\mathrm{dep}$ are comparable, the net effect on metallicity is small and the $Z_\mathrm{gas}$--$\Sigma_\mathrm{SFR}$ relation becomes weak or flat.

\wl\ also explain this (anti-)correlation from a theoretical point of view. Their model attributes the positive case (in $\sim 100$\,pc scales) to time-variable SFE acting against a relatively steady gas inflow rate, and the negative case (at galactic scales) to strongly time-variable inflow acting against nearly constant SFE. Our MAUVE--MUSE galaxies, however, show the coexistence of both positive and negative correlations already at $\sim 100$\,pc scales, and our regulator analysis indicates that the sign is set locally by the \emph{relative amplitude} of variations in $\Sigma_\mathrm{SFR}$ and the gas-supply-rate surface density ($\Sigma_\Phi$), rather than by scale alone. If the spaxel-to-spaxel changes in $\Sigma_\mathrm{SFR}$ are interpreted as being driven primarily by fluctuations in the SFE of individual GMCs, while $\Sigma_\Phi$ is taken as the local analogue of an inflow rate, then  \autoref{eq:deltaZ_case2_final} is fully consistent with the \wl\ framework: positive (negative) correlations emerge where SFE- or SFR-driven (GSR-driven) variability dominates, and both regimes can naturally coexist within the same galaxy at fixed spatial resolution.

On $\sim$kpc scales, \citet{sanchezmenguiano19} used MaNGA spaxels and the \marino\ strong-line calibration to show that the sign and strength of the local $\Delta Z_{\rm gas}$--$\Delta\Sigma_{\rm SFR}$ relation are primarily set by a galaxy's average metallicity: metal-poor systems exhibit the steepest negative slopes, whereas metal-rich systems show a much weaker (or near-zero) trend. They report an inversion at $12+\log_{10}({\rm O/H})\simeq 8.5$, which corresponds to $\log_{10}(\Sigma_\ast)\simeq 7.5$ using our best-fitting rMZR parameters in \autoref{eq:rMZR_fit}, and is therefore consistent with the inversion point we identify. While our limited sample size ($N=14$) is insufficient to show an obvious monotonic dependence of the reversal strength on galaxy-averaged metallicity, we note that NGC~4396, which is our most strongly anti-correlated case in \autoref{fig:spearman}, also has the lowest integrated oxygen abundance over the \textbf{HII}-selected spaxels (by $\sim$0.15\,dex relative to the rest, for O3N2-based abundances). In \citet{sanchezmenguiano19}'s interpretation, this behaviour reflects a shift in the origin of the gas fueling star formation: in metal-poor galaxies, localized star-formation enhancements are associated with external accretion of metal-poor gas (strong dilution, hence negative slopes), whereas in metal-rich galaxies the fueling is increasingly dominated by recycled/internal gas (reduced dilution, allowing enrichment to dominate and the relation to flatten or become positive). In the language of our gas-regulator model, these two regimes map naturally onto which term controls the short-timescale departures in $Z_{\rm gas}$. If fluctuations in the gas-supply term dominate, the response is dilution-driven and $\Delta Z_{\rm gas}$ anti-correlates with $\Delta\Sigma_{\rm SFR}$; if instead it is tied to the star-formation/feedback cycle (i.e. $\tau_{\rm dep}$), enrichment can dominate and the relation becomes weak or positive. 

Because our galaxies reside in the Virgo Cluster, environmental processes that perturb gas accretion, stripping, and recycling may modulate the local coupling between $Z_{\rm gas}$ and $\Sigma_\mathrm{SFR}$, plausibly enhancing the role of gas-supply regulation relative to enrichment \citep{cortese21}. However, the mass-dependent inversion we observe is not simply a radial-selection effect, because similar behaviour is seen when comparing regions at comparable galactocentric distances across different galaxies. Therefore, we argue that the anti-correlation in the low-$\Sigma_*$ regime is not purely environmentally driven, but instead we treat the environment as a plausible contributing factor rather than the primary explanation. 

Finally, we note that converting dust-corrected H$\alpha$ to $\Sigma_{\rm SFR}$ assumes that star formation has been approximately constant over the H$\alpha$ tracer timescale (a few–$\sim$10 Myr); at $\sim$100\,pc this assumption can break down, introducing stochastic (burstiness) fluctuations in the inferred $\Sigma_{\rm SFR}$. In practice, this effect enters our framework as an additional contribution to $\Delta\log_{10}(\Sigma_{\rm SFR})$ in \autoref{eq:deltaZ_case2_final}, i.e. it is absorbed into the ``SFR-driven variability'' term ($\delta\log_{10}(\Sigma_\mathrm{SFR}(t))$). If this extra variance is largely uncorrelated with $Z_{\rm gas}$, it will primarily inflate the scatter and dilute the observed $\Delta Z_{\rm gas}$--$\Delta\Sigma_{\rm SFR}$ trend in regimes where the intrinsic relation is positive; conversely, in regimes where the observed relation is dilution-driven and negative, the same additional excursions in $\Delta\log_{10}(\Sigma_{\rm SFR})$ can make the anti-correlation appear more pronounced. This raises the possibility that part of the inferred inversion could be influenced by H$\alpha$-based $\Sigma_{\rm SFR}$ stochasticity; explicitly accounting for such fluctuations would primarily reduce the apparent significance/contrast of the sign-change pattern in \autoref{fig:spearman}. However, this cannot be tested with the current data.

\subsection{From Spatially-resolved to Integrated Relations}
\label{sec:extrapolation}

Although this theoretical framework presented so far is written entirely in terms of surface densities, the underlying regulator equations have exactly the same functional form if one replaces $(\Sigma_\mathrm{g},\Sigma_*,\Sigma_\mathrm{SFR},\Sigma_\Phi)$ by their galaxy–integrated counterparts $(M_\mathrm{g},M_*,\mathrm{SFR},\Phi)$ and the associated timescales by effective global depletion and inflow timescales. In this sense, \autoref{eq:deltaZ_case2_final} can be read as a scale–free statement: at fixed $M_*$, the sign of the $Z_\mathrm{gas}$–SFR (anti-)correlation is determined by the comparison of the relative fluctuation amplitudes in SFR and in the inflow rate, or equivalently by whether variations in depletion or inflow timescale dominate. If global fluctuations in SFR (or SFE) are larger than those in inflow, one expects a positive $Z_\mathrm{gas}$–SFR correlation at fixed $M_*$; if stochasticity in inflow dominates, dilution produces an anti-correlation. This picture provides a natural physical interpretation for the integrated fundamental metallicity plane/surface found in large samples, where the SFR dependence of metallicity is negative at low stellar mass and commonly observed but weakens and may even change sign at the high-mass end \citep[e.g.,][]{mannucci10,yates12,zahid13,curti20}. In our view, these global trends reflect the same basic competition between star formation and gas inflow encapsulated in \autoref{eq:deltaZ_case2_final}, with low-mass galaxies residing in a regime where inflow-driven variability is more significant, and massive systems shifting toward a regime where SFR- or SFE-driven variability becomes comparatively stronger.

Within this picture, the apparent prevalence of anti-correlations in both resolved and integrated studies can be understood, at least in part, as a consequence of how different surveys sample the $M_*$ and/or $\Sigma_*$ parameter space. Global FMR analyses are dominated by galaxies with $M_* \lesssim 10^{10},M_\odot$, while more massive systems with $M_* \gtrsim 10^{10.5}\,M_\odot$ are rarer and span a narrower range of specific SFR, so the low-mass, inflow-dominated regime contributes a larger fraction of the statistical weight. An analogous effect arises in spatially resolved work: at kpc-scale resolution, some surveys (e.g., MaNGA and MAGPI) preferentially sample spaxels with $\log_{10}(\Sigma_*/M_\odot\,\mathrm{kpc}^{-2}) \lesssim 7.8$ and therefore see predominantly anti-correlated or flat $Z_\mathrm{gas}$–$\Sigma_\mathrm{SFR}$ behaviour \citep{baker23,koller24}, whereas at $\sim 100$\,pc resolution our MAUVE–-MUSE map, and possibly MAD data in \wl, contain a much larger fraction of spaxels with $\log_{10}(\Sigma_*/M_\odot\,\mathrm{kpc}^{-2}) \gtrsim 8$, so the positive branch dominates the ensemble trends. In both the global and resolved cases, the sign flip emerges only when one samples across the transition region (around $\log_{10}(\Sigma_*/M_\odot\,\mathrm{kpc}^{-2})\sim 7.5-8.0$ locally and $\log_{10}(M_*/M_\odot) \sim 10.0-10.5$ globally), and the balance between correlated and anti-correlated regimes is then controlled by the relative number of galaxies/spaxels on either side of this inversion. Furthermore, the more frequent detection of anti-correlation, as well as the weaker and less clear (anti-)correlation signal on global scales than on local scales, may simply be a result of an averaging effect, whereby integrating over many spaxels blends regions together with opposite local behaviours \citeg{wang&lilly21}. However, (i) what sets the precise location of these transition scales, (ii) whether the local ($\Sigma_*$) and global ($M_*$) inversions are causally linked or merely coincident, and (iii) which physical conditions change across them (e.g., gas fraction, disc stability, outflow efficiency, or environmental processes), remain open questions that we defer to future work. 





\section{Robustness of Our Findings}
\label{sec:discussions}
 
\begin{figure*}
    \centering
    \includegraphics[width=\textwidth]{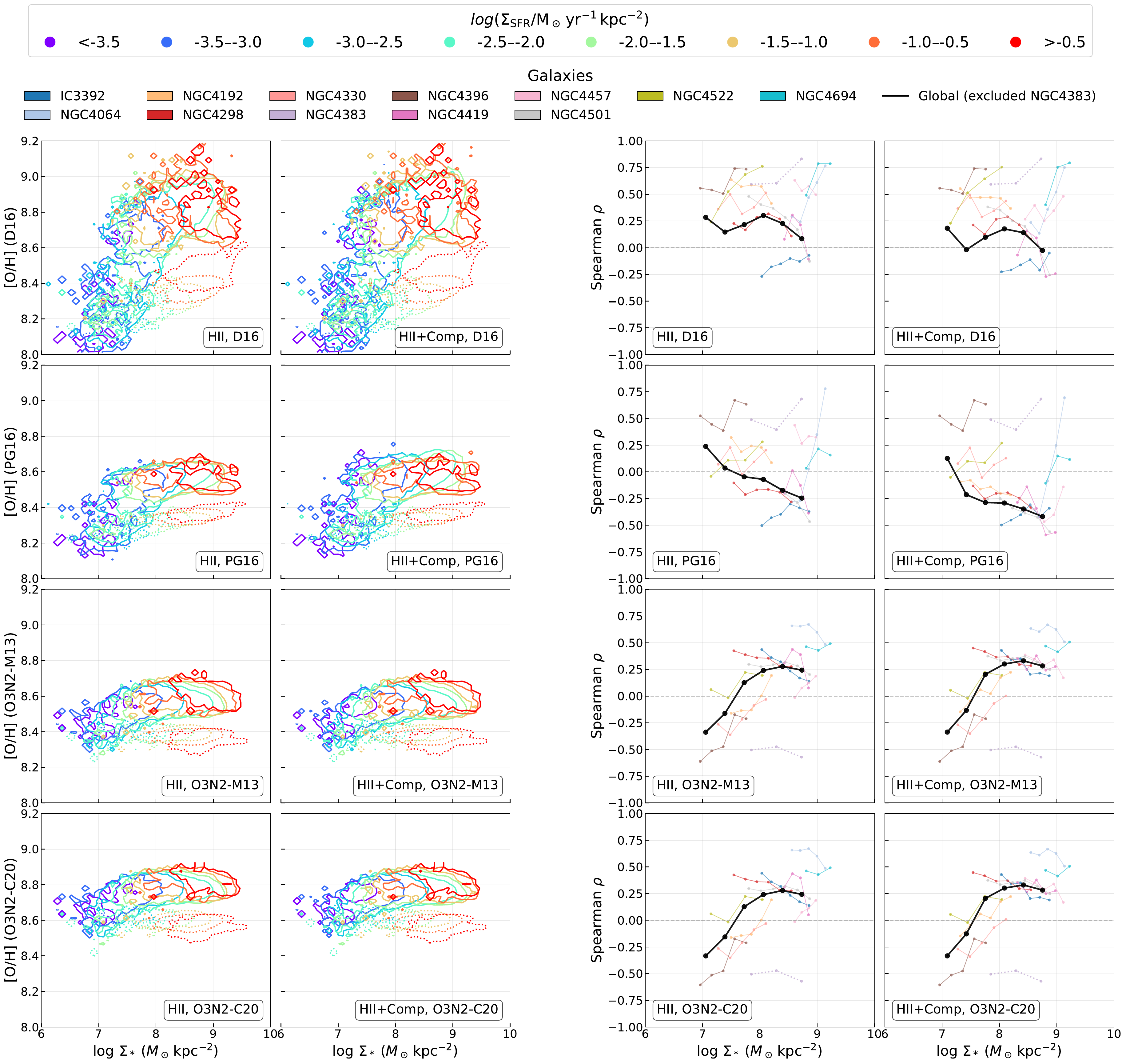}
    \caption{Comparison of the rMZR (left two columns) and the mass-dependent Spearman rank correlation coefficient $\rho$ between $\Delta[{\rm O/H}]$ and $\Delta\log_{10}\Sigma_{\rm SFR}$ (right two columns) for four gas-phase metallicity calibrations: \dopita\ (top row), \pg\ (second row), \marino\ (third row), and \curti\ (bottom row). Columns 1 and 3 use only \textbf{HII}-classified star-forming spaxels, whereas Columns 2 and 4 use the more inclusive \textbf{HII+Comp} selection (see the \bpt\ demarcation in \autoref{fig:bpt}). The rMZR panels are colour-coded by $\log_{10}\Sigma_{\rm SFR}$ as in \autoref{fig:main}. The Spearman panels show $\rho(\Sigma_*)$ measured in discrete $\Sigma_*$ bins for individual galaxies (coloured curves) and for the ensemble sample (black curve; excluding NGC~4383), following \autoref{fig:spearman}. The parameter spaces occupied by NGC~4383 are indicated by dotted contours/curves in each corresponding panel.}
    \label{fig:4x4}
\end{figure*}

In this section, we test whether the $Z_{\rm gas}$--$\Sigma_{\rm SFR}$ (anti-)correlation is robust to three potential systematics: (i) different strong-line abundance prescriptions (\autoref{sec:discussion_metallicity}), (ii) alternative selections of star-forming spaxels (\autoref{sec:discussion_sf}), and (iii) the impact of the metallicity outlier NGC~4383 (\autoref{sec:discussion_4383}). A compact overview is shown in \autoref{fig:4x4}. The broader purpose of these checks is to emphasise that, although calibration choice, excitation mixing, environment, and mass-dependent sampling can modulate the apparent scatter and the detailed slope/normalization of rMZR in individual regimes, they do not alter the underlying physical interpretation implied by \autoref{eq:deltaZ_case2_final}, namely that the sign of the local residual correlation is set by the competition between the relative fluctuations of star formation and gas supply, $\delta\log_{10}\Sigma_{\rm SFR}$ versus $\delta\log_{10}\Sigma_\Phi$, or equivalently by the balance between gas supply and gas depletion timescales,$\delta\log_{10}\tau_\Phi$ and $\delta\log_{10}\tau_{\rm dep}$.

\subsection{Sensitivity to Metallicity Calibration}
\label{sec:discussion_metallicity}

A well-known limitation of any resolved metallicity analysis is that strong-line prescriptions are not anchored to a unique absolute scale. Different calibrations rely on different reference samples and methodologies—direct electron-temperature ($T_e$) measurements, photoionization model grids, or hybrid/stacked approaches—and they respond differently to variations in excitation, ionization parameter, DIG, and N/O abundance. Consequently, systematic offsets of order $\sim 0.1$--$0.3$ dex (and in extreme cases larger by up to $\sim 0.5$ dex) between calibrations applied to the same spectra are well documented, and can alter both the normalization and the slopes of resolved relations \citep[e.g.,][]{kewley&ellison08,kewley19,maiolino&mannucci19,scudder21}. These long-standing uncertainties motivate our use of four widely adopted diagnostics spanning empirical $T_e$-calibrated and model-based families: \dopita, \pg, \marino, and \curti.

The first column of \autoref{fig:4x4} shows that, despite calibration-dependent zero-points, the overall rMZR morphology is broadly stable across all prescriptions: $Z_{\rm gas}$ increases steeply with $\Sigma_*$ at low surface densities and flattens toward a saturation-like regime at high $\Sigma_*$, in line with previous resolved surveys. 
However, the secondary dependency on $\Sigma_{\rm SFR}$ is not equally coherent for every diagnostic. For both O3N2-based calibrations, the $\Sigma_{\rm SFR}$-stratified contours display a clear and systematic reversal with $\Sigma_*$: at low $\Sigma_*$, higher $\Sigma_{\rm SFR}$ preferentially occupies lower $Z_{\rm gas}$, while at high $\Sigma_*$ the ordering flips such that higher $\Sigma_{\rm SFR}$ corresponds to higher $Z_{\rm gas}$. This behaviour is reflected in the ensemble Spearman trends (third column), which show a coherent transition from $\rho<0$ at low $\Sigma_*$ to $\rho>0$ at intermediate/high $\Sigma_*$ for both O3N2 prescriptions (with galaxy-to-galaxy scatter but a consistent global trend). By contrast, the \dopita\ and \pg\ prescriptions exhibit substantially weaker and more heterogeneous $\Sigma_{\rm SFR}$ stratification in the H\,\textsc{ii}-only sample. The \dopita\ ensemble $\rho$ trend remains mildly positive over most of the $\Sigma_*$ range; while \pg\ shows a stronger dependence on $\Sigma_*$ in the ensemble statistic, but with opposite-sign behaviour compared to \marino\ and \curti. Therefore, while the rMZR curvature itself is stable, the inferred $Z_{\rm gas}$--$\Sigma_{\rm SFR}$ coupling (and any apparent sign change) is diagnostic-dependent.

The comparison also reveals systematic differences between calibration families. In our sample, indicators tied more closely to $T_e$-based anchors (e.g.\ \pg\ and both O3N2 calibrations) show tighter rMZR trends with metallicity dispersions of $\sim0.1-0.15$ dex at fixed $\Sigma_*$ and $\Sigma_{\rm SFR}$, whereas the model-based \dopita\ prescription produces visibly larger scatter ($\sim0.2-0.25$ dex) and spans a broader metallicity range ($\sim0.4-0.5$ dex wider) across the same parameter space. This behaviour is consistent with PHANGS–MUSE \citep{kreckel19}, which shows that calibrations tied to photoionization models or to a small number of line ratios are more susceptible to local variations in ionization parameter and abundance ratios, leading to increased spaxel-to-spaxel dispersion and systematic offsets relative to $T_e$-anchored scales. Within the O3N2 family itself, the two prescriptions are highly consistent in their trends; the main difference is a nearly uniform offset, with \curti\ yielding metallicities systematically higher than \marino\ by $\sim 0.1$--$0.2$ dex.

While both O3N2-based calibrations recover a similar inversion, the presence and coherence of the sign change are not universal across all prescriptions. The \dopita\ Spearman trends are dominated by positive $\rho$ over most of the $\Sigma_*$ range, mirroring the predominantly positive resolved correlation reported by \wl\ on $\sim 100$ pc scales. This suggests that the \emph{choice of calibration} can influence whether a given data set appears dominated by correlated or anti-correlated regimes, even when the underlying spaxel population is the same. Moreover, at the high-$\Sigma_*$ end, the O3N2 prescriptions favour a positive residual coupling, whereas \dopita\ and \pg\ lean toward weaker or even negative behaviour. Similar calibration-dependent qualitative differences have been noted in other MUSE-based surveys, where O3N2- and N2-type indicators can yield different resolved slopes from S- or N2S2-based prescriptions \citep[e.g.,][]{kreckel19}. We therefore caution that the detailed mass dependence of the $Z_{\rm gas}$--$\Sigma_{\rm SFR}$ relation can diminish or even reverse under different metallicity calibrations. 

In general, both O3N2 prescriptions are also the ones that most cleanly delineate the specific behaviour of interest here: they show a coherent transition from an anti-correlation between $Z_{\rm gas}$ and $\Sigma_{\rm SFR}$ at low $\Sigma_*$ to a positive coupling at higher $\Sigma_*$. In our resolved sample, this inversion occurs at $\log_{10}(\Sigma_*/M_\odot\,{\rm kpc}^{-2})\approx 7.5$--$8.0$. From a physical point of view, such a transition is qualitatively expected in gas-regulator frameworks: in low-$\Sigma_*$ (typically lower-metallicity) regimes, fluctuations in gas supply can dominate the short-timescale response and produce dilution-driven anti-correlations, whereas in higher-$\Sigma_*$ regimes the response can be increasingly enrichment- or consumption-dominated, yielding a positive residual coupling. This qualitative pattern is consistent with the global, integrated picture established using SDSS data, whose oxygen abundances derived by both strong line diagnostics and grids of photoionization models reveal a similar reversal, negative at low masses and positive at high masses, with a transition point around a characteristic mass of $\log_{10}(M_*/M_\odot)\approx10.0-10.5$ (see Figure 1 of \citealt{yates12} and Figure 4 of \citealt{zahid13}). Although the signal is not strong, \citet{curti20} also present such a reversal of $Z_{\rm gas}$--$\Sigma_{\rm SFR}$ correlation at the high-mass end in their Figure 6 and 7. Similarly, such a transition in the global scenario is also probed in Evolution and Assembly of GaLaxies and their Environments (EAGLE) suite of cosmological hydrodynamical simulations \citep{trayford19}.

In addition, we emphasize that the inferred sign structure may be calibration-dependent. For example, \dopita\ implies that the residual coupling is predominantly positive across most of the $\Sigma_*$ range, which would suggest enrichment-dominated behaviour even in the lowest-$\Sigma_*$ regime probed here. Given that our galaxy sample spans a limited range in global stellar mass, we cannot rule out that more low-mass systems would strengthen the dilution-dominated branch; moreover, \dopita\ relies on ratios involving [\ion{N}{ii}] and [\ion{S}{ii}], which can be particularly sensitive to variations in N/O and to DIG/shock contributions at $\sim$100~pc resolution, potentially biasing $\rho(\Sigma_*)$ toward positive values and/or increasing the scatter. In contrast, \pg\ yields a less coherent $\rho(\Sigma_*)$ behaviour and can imply comparatively stronger anti-correlation at high $\Sigma_*$, which is difficult to reconcile with the integrated reversal and basic regulator expectations, suggesting that systematic sensitivity to local excitation/line-ratio mixing may dominate in this calibration. We therefore adopt the O3N2 prescriptions as our fiducial description of the inversion while treating \dopita\ and \pg\ primarily as systematic cross-checks on calibration-dependent behaviour.

\subsection{Sensitivity to the Star-forming Spaxel Definition}
\label{sec:discussion_sf}

Our baseline analysis relies on a conservative \textbf{HII}-only mask, in which spaxels are required to fall within the \textbf{HII} locus of both the [\ion{N}{ii}] and [\ion{S}{ii}] \bpt\ diagrams with robust line detections and uncertainties that remain fully inside the same excitation class, together with the $\mathrm{EW(H}\alpha) > 6$\,\AA\ and $\sigma(\mathrm{H}\alpha) < 45\,\mathrm{km\,s^{-1}}$ criterion. This stringent selection is designed to minimise contamination from shocks, AGN-like excitation, and DIG, all of which can bias strong-line metallicity indicators and H$\alpha$-based $\Sigma_\mathrm{SFR}$ estimates. However, MAUVE--MUSE also provides a more inclusive star-forming mask that merges \textbf{HII} and composite regions (\textbf{HII+Comp}), motivated by the expectation that composite spaxels are still largely powered by O/B stars' photoionization but may include a non-negligible contribution from DIG or harder radiation fields, to which we likewise apply the $\mathrm{EW(H}\alpha) > 6$\,\AA\ and $\sigma(\mathrm{H}\alpha) < 45\,\mathrm{km\,s^{-1}}$ selection. We therefore repeat the full rMZR and offset-correlation analysis using \textbf{HII+Comp} spaxels to assess the sensitivity of our conclusions to excitation masking.

The comparison is summarised in \autoref{fig:4x4} as well, where Columns 1 and 3 use the conservative \textbf{HII} selection, and Columns 2 and 4 adopt \textbf{HII+Comp}. Including composite spaxels expands the sampled parameter space, especially toward high $\Sigma_*$ where inner discs and circumnuclear regions are more likely to host mixed excitation and more complex physics. For the two O3N2-based prescriptions (\marino\ and \curti), both the rMZR colour ordering and the mass-dependent Spearman trends are essentially unchanged when composites are added. The ensemble $\rho(\Sigma_*)$ curves retain the same negative-to-positive transition at $\log_{10}(\Sigma_*/M_\odot\,\mathrm{kpc}^{-2}) \sim 7.5$--8.0, and the scatter around the median rMZR locus increases only marginally. This stability indicates that our main result is not driven by a narrow excitation-class definition, and that O3N2 indicators are comparatively resilient to moderate DIG/composite admixture in the MAUVE--MUSE data.

By contrast, the N2S2- and S-calibrations show a stronger response to including composites. In particular, the \pg\ rMZR exhibits a noticeable increase in scatter at the high-$\Sigma_*$ end under the \textbf{HII+Comp} mask, consistent with the known sensitivity of S2- and N2-based ratios to harder ionizing spectra and DIG-enhanced low-ionization lines. The corresponding $\rho(\Sigma_*)$ trends also shift: for \dopita\ the ensemble Spearman curve becomes dominated by weak or negative correlations over a broader $\Sigma_*$ range, and some individual galaxies show reduced significance or a delayed zero-crossing; while for \pg\ the ensemble Spearman trend becomes completely negative without an inversion point anymore. These changes underline that the N2S2- and S-calibrations (i.e. \dopita\ and \pg\ here) leveraging [\ion{N}{ii}] and [\ion{S}{ii}] can be more susceptible to excitation mixing, and they reinforce the need to treat excitation masking and calibration choice as coupled systematics when interpreting resolved $Z_\mathrm{gas}$--$\Sigma_\mathrm{SFR}$ relations.

Crucially, despite these quantitative shifts, the qualitative inversion of the residual $Z_\mathrm{gas}$--$\Sigma_\mathrm{SFR}$ coupling persists in the ensemble sample for both \marino\ and \curti. The low-$\Sigma_*$ anti-correlation branch and the high-$\Sigma_*$ positive branch remain present, demonstrating that the sign flip is not an artefact of excluding composite spaxels when using O3N2 prescriptions. 

We also test whether our main result depends on deriving $\Sigma_*$, $\Sigma_\mathrm{SFR}$, and $Z_\mathrm{gas}$ at the native $\sim$100\,pc scale. While we acknowledge that nebular diagnostics require care when approaching the scale of individual \ion{H}{ii} regions, recent MUSE studies routinely perform H$\alpha$-based SFR and strong-line metallicity measurements at these resolutions, provided that the analysis is restricted to \ion{H}{ii}-region-dominated emission and DIG contamination is minimised \citep{errozferrer19,wang&lilly21,belfiore23a,brazzini24,easeman24}. Moreover, this concern is unlikely to dominate our analysis, since most \ion{H}{ii} regions are still unresolved or only marginally resolved at $\sim$100\,pc resolution \citep{kreckel19,santoro22,groves23}. Therefore, although the largest and brightest complexes may be partially resolved, the bulk of the population at these resolutions is still appropriately treated as individual \ion{H}{ii} regions or only marginally resolved nebulae, such that the observed fluctuations in $\Sigma_\mathrm{SFR}$ and $Z_\mathrm{gas}$ predominantly trace variations among \ion{H}{ii} regions rather than internal sub-\ion{H}{ii}-region structure. Additionally, we can confirm that our main findings do not qualitatively change if we perform a similar analysis at 500 parsec resolution.

Finally, we note that many galaxies in our sample are early spirals with substantial bulge components. In this context, the high-$\Sigma_*$ regime may not simply trace the ordinary star formation in disc, but could also reflect the influence of bulge-dominated environments, where star formation is known to be suppressed relative to discs at fixed mass \citep{mendezabreu19}. Structural effects such as bulge growth or morphological quenching may therefore contribute to the distinct behaviour seen at high $\Sigma_*$ regions.

\subsection{NGC4383 -- the Metallicity Outlier}
\label{sec:discussion_4383}

NGC~4383 is a clear metallicity outlier within the MAUVE--MUSE sample in the spatially resolved scenario. To illustrate its behaviour, we reintroduce its spaxels in \autoref{fig:4x4} as dotted contours (rMZR panels) and dotted light-mauve curves (Spearman panels), while keeping the ensemble Spearman trend (black curve) computed from the other 13 galaxies only. From the rMZR comparison (first two columns), NGC~4383 remains systematically offset to lower $Z_{\rm gas}$ at fixed $\Sigma_*$ by at least 0.2 dex under all four metallicity calibrations. The persistence of this depression across prescriptions and excitation selections indicates that it is not produced by calibration-specific systematics or by the choice of star-forming mask, motivating our designation of NGC~4383 as a metallicity outlier.

The Spearman trend of NGC~4383 adds a second, more striking peculiarity. Unlike the other galaxies, NGC~4383 shows a mass-dependent correlation pattern that is nearly the mirror image of the ensemble relation: wherever the combined MAUVE--MUSE behaviour is positive (negative), the NGC~4383 curve tends to be negative (positive), especially in O3N2 indicators. In other words, the sign of the residual $Z_{\rm gas}$--$\Sigma_{\rm SFR}$ coupling in NGC~4383 appears to reverse relative to the dominant local regime traced by the rest of the sample. If included in the stack, this anti-aligned behaviour would disproportionately populate the low-$Z_{\rm gas}$ tail and partially cancel the mass-ordered trend, artificially steepening the low-$\Sigma_*$ rMZR branch and shifting $\rho(\Sigma_*)$ toward the non-correlation case. This exclusion is therefore not motivated by statistical convenience, but by the fact that NGC~4383 represents a physically distinct system whose resolved ionized-gas properties are not comparable to those of the remaining MAUVE--MUSE disc sample. Such abnormal behaviour of NGC~4383 in \autoref{fig:4x4} under the O3N2 indicators implies that, in contrast to the other 13 MAUVE--MUSE galaxies, NGC~4383 is dominated by GSR-driven variability in the $\log_{10}(\Sigma_*/M_\odot\,\mathrm{kpc}^{-2})\gtrsim 7.8$ regime.

A plausible physical explanation for both its depressed metallicity scale and its opposite-sign residual coupling is a substantial reservoir of (re-)accreted, relatively metal-poor gas that enhances gas-supply variability and dilutes $Z_{\rm gas}$ at fixed $\Sigma_*$. This interpretation is consistent with the kinematic anomalies reported by \citet{chung09} and with the view that NGC~4383 has recently acquired a large gas reservoir \citep{cortese26}. 

Moreover, \citet{watts24} show that NGC~4383 hosts a $\sim$6\,kpc bipolar ionized outflow with complex, clumpy structure and an average outflow velocity of $\sim 210\,\mathrm{km\,s^{-1}}$, launched from one of the most H\,\textsc{i}-rich discs in Virgo. They further find that the outflowing gas is marginally more metal-rich than the disc but less enriched than the central starburst, consistent with mixing between ejected and entrained material. On the other hand, \citet{cortese26} reported cold neutral gas associated with the outflow and argued that NGC~4383 is likely experiencing a multiphase fountain flow. Together, these results strongly suggest that the local ionized-gas properties of NGC~4383 are shaped by a disturbed, non-disc gas cycle rather than by ordinary in-situ disc \ion{H}{ii} regions alone. NGC~4383 therefore provides an existence proof that the commonly observed rMZR and the local $Z_{\rm gas}$--$\Sigma_{\rm SFR}$ coupling can appear to break down when the ionized emission is strongly affected by non-disc gas (e.g., outflows/extraplanar components) and/or when the system is subject to strong, recent gas-supply perturbations. Intriguingly, NGC~4383 is also qualitatively consistent with the spatially resolved gas-regulator framework developed in this paper, which predicts that an anti-correlation between star formation and metallicity should arise when fluctuations in the gas supply rate dominate over fluctuations in the star formation efficiency, as is likely the case in this galaxy.

\section{Conclusions}
\label{sec:conclusions}

In this paper, we have analysed 14 Virgo spiral galaxies from the MAUVE--MUSE second internal release to revisit the resolved mass--metallicity relation and its secondary dependence on SFR surface density on $\sim 100$\,pc scales. By combining high-quality IFS data with a spatially resolved regulator framework, we have characterised where and why the local gas-phase metallicity to SFR surface density ($Z_\mathrm{gas}$--$\Sigma_\mathrm{SFR}$) relation (and also its residual coupling) changes sign. Our main conclusions are as follows.

\begin{enumerate}
    \item Our MAUVE--MUSE galaxies follow a well-defined resolved mass--metallicity relation that is consistent with previous IFS surveys. Using the \marino\ calibration as our fiducial metallicity scale, the resolved mass metallicity relation (rMZR) rises steeply at low stellar mass surface density ($\Sigma_*$) and flattens toward high $\Sigma_*$, in line with earlier work on kpc-scale data. At fixed $\Sigma_*$, low-$\Sigma_*$ regions tend to show lower $Z_\mathrm{gas}$ at higher $\Sigma_\mathrm{SFR}$, whereas high-$\Sigma_*$ regions show the opposite ordering (see \autoref{fig:main}).

    \item After removing the primary rMZR and resolved star formation main sequence (rSFMS) trends, the residual coupling between metallicity and SFR surface density exhibits a robust mass-dependent inversion. The Spearman rank correlation coefficient between $\Delta[{\rm O/H}]$ and $\Delta\log_{10}\Sigma_\mathrm{SFR}$ transitions smoothly from $\rho<0$ at $\log_{10}(\Sigma_*/M_\odot\,\mathrm{kpc}^{-2})\lesssim 7.5$ to $\rho>0$ at $\log_{10}(\Sigma_*/M_\odot\,\mathrm{kpc}^{-2})\gtrsim 8.0$, with a zero-crossing in between (see \autoref{fig:spearman}). Notably, NGC~4192 spans a broad $\Sigma_*$ range that straddles the inversion point and exhibits a clear internal sign reversal, rather than simply populating one branch or the other. This suggests that the transition is driven by an intrinsic physical mechanism operating within individual galaxies, rather than arising as an artifact of stacking different galaxy populations. Local Moran-like correlation maps further show that, once $\Sigma_*$ trends are removed, correlated and anti-correlated \ion{H}{ii} regions can coexist within the same galaxy, and the familiar mass-dependent behaviour emerges only as an ensemble property.

    \item To interpret these patterns, we formulated a spatially resolved gas--regulator model in surface densities and identified an ``instantaneous target metallicity'' ($Z^\dagger$) that drives the evolution of the observed real-time metallicity ($Z$). Analysis around metallicity extrema and toward/away from them leads to a simple condition (see \autoref{eq:deltaZ_case2_final}): at fixed $\Sigma_*$, the sign of the local $Z_\mathrm{gas}$--$\Sigma_\mathrm{SFR}$ (anti-)correlation is controlled by the relative fluctuation amplitudes of star formation and gas supply. Positive correlations arise where variability of SFR or star formation efficiency dominates (short, strongly varying depletion times at nearly steady gas supply), while negative correlations arise where variability of gas supply rate dominates and dilution drives $Z_\mathrm{gas}$ opposite to $\Sigma_\mathrm{SFR}$. Our MAUVE--MUSE trends are therefore naturally understood as the outcome of a competition between SFR-driven and GSR-driven fluctuations that vary systematically with $\Sigma_*$. The physical drivers of these variabilities will be explored in future work.

    \item We argued that the same gas-regulator physics can be extrapolated from spatially resolved to integrated galaxy-wide relations (\autoref{sec:extrapolation}). Replacing surface densities by their global analogues, \autoref{eq:deltaZ_case2_final} implies that the sign of the integrated $Z_\mathrm{gas}$--SFR dependence at fixed $M_*$ is likewise set by whether fluctuations in SFR (or SFE) outweigh those in the inflow rate. In this view, the negative SFR dependence of metallicity commonly seen at low stellar masses and the weaker or reversed dependence at the high-mass end in global FMR analyses are two manifestations of the same underlying competition between depletion and inflow timescales. The apparent prevalence of anti-correlation, and the weaker and less clear direct correlation signal on global scales than on local scales, can be understood as a consequence of parameter-space sampling and spatial averaging: many surveys predominantly probe galaxies or spaxels on the low-mass, inflow-dominated side of the inversion and then integrate over regions with mixed local behaviour.

    \item Finally, we tested the robustness of these conclusions against three key systematics (see \autoref{fig:4x4}). First, we tested four widely used metallicity diagnostics (\dopita, \pg, \marino\ and \curti) and find that, although the resolved metallicity scaling relations are qualitatively similar across diagnostics, the presence and sign of the inferred $Z_\mathrm{gas}$--$\Sigma_\mathrm{SFR}$ inversion are calibration-dependent. The clearest and most robust mass-dependent inversion is recovered for the two O3N2-based calibrations (\marino\ and \curti). By contrast, the N2S2- and S-based diagnostics (\dopita\ and \pg) respond more strongly to excitation mixing and show substantially different behaviour: under the \pg\ prescription the coupling becomes predominantly negative across the full $\Sigma_*$ range, while \dopita\ shows no clear monotonic correlation, with the stacking Spearman trend fluctuating around zero. Second, relaxing the star-forming mask from a conservative \textbf{HII}-only selection to an inclusive \textbf{HII+Comp} mask increases the scatter and modulates the detailed $\rho(\Sigma_*)$ trends, with the strongest changes again occurring for the N2S2- and S-based diagnostics. Third, we showed that NGC~4383 is a genuine metallicity outlier: it lies $\sim 0.1$--$0.2$\,dex below the rMZR locus and exhibits an almost mirror-image residual correlation pattern relative to the stack.

\end{enumerate}

In summary, these results suggest that the spatially resolved $Z_\mathrm{gas}$--$\Sigma_\mathrm{SFR}$ relation is not governed by a single universal trend, but by a balance between star-formation-driven and gas-supply-driven variability that depends on local mass surface density, environment, and feedback. Our MAUVE--MUSE data show that both correlated and anti-correlated regimes naturally coexist within the same discs at $\sim 100$\,pc resolution, and that the familiar mass-dependent inversion emerges as a population-level imprint of this underlying competition. At the same time, the detailed presence and sign of the inversion are not calibration-independent: the effect is recovered most clearly for the O3N2-based metallicity prescriptions, while N2S2- and S-based diagnostics are more sensitive to excitation mixing and yield less consistent behaviour. Future work combining MAUVE--MUSE data with resolved \ion{H}{i} and CO observations, as well as higher-redshift IFS surveys will be crucial for constraining gas supply rate surface density ($\Sigma_\Phi$) and its timescale more directly, and for testing whether the inversion scales identified here are universal or themselves evolve with redshift, environment, and galaxy assembly history.

\section*{Software}

This research made use of \texttt{astropy} \citep[\url{https://www.astropy.org/}]{astropy22}, \texttt{numpy} \citep[\url{https://numpy.org}]{numpy}, \texttt{matplotlib} \citep[\url{https://matplotlib.org/}]{matplotlib}, \texttt{scipy} \citep[\url{https://scipy.org/}]{scipy}, \texttt{speclite} \citep[\url{https://speclite.readthedocs.io/en/latest/}]{kirkby23} and \texttt{nGIST} \citep[\url{https://geckos-survey.github.io/gist-documentation/}]{fraser-mckelvie25}.

\section*{Acknowledgements}
The authors are grateful to the referee, Sebastian Sanchez, who provided valuable and insightful comments that improved the quality of this manuscript. This work is carried out as part of the MAUVE collaboration (\href{https://mauve.icrar.org/}{https://mauve.icrar.org/}) and is based on observations collected at the ESO under ESO programme(s) 105.208Y and 110.244E. The authors acknowledge the use of the Canadian Advanced Network for Astronomy Research (CANFAR) Science Platform operated by the Canadian Astronomy Data Center (CADC) and the Digital Research Alliance of Canada (DRAC), with support from the National Research Council of Canada (NRC), the Canadian Space Agency (CSA), CANARIE, and the Canadian Foundation for Innovation (CFI). RH thanks Christine D. Wilson for valuable comments on the manuscript, and Enci Wang, Yunhe Li, Sriram Sankar, Ajay Dev, Amy Attwater, Shinna Kim, Turman Tapia Mora, Tim Gao, Yao Yao, and Yusen Li for helpful discussions. LC acknowledges support from the Australian Research Council Discovery Project funding scheme (DP210100337). LJMD acknowledges support from the Australian Research Council Future Fellowship funding scheme (FT200100055). AR acknowledges the support by the Australian Research Council Centre of Excellence in Optical Microcombs for Breakthrough Science (project number CE230100006), with funding from the Australian Government. VV acknowledges support from the Comité ESO Mixto 2024 and from the ANID BASAL project FB210003. This research has made extensive use of NASA’s Astrophysics Data System Bibliographic Services.

\section*{Data Availability}
The full set of MAUVE--MUSE datacubes and associated \texttt{nGIST} value-added products will be released as part of the main public data release once the survey is completed (anticipated in late 2027). Requests for access prior to the public release should be directed to the corresponding author. Readers interested in joining the MAUVE collaboration and accessing the full dataset are encouraged to consult the \href{https://mauve.icrar.org/team.html}{team} page on the MAUVE survey website. The analysis pipeline used in this work is publicly available at \url{https://github.com/Rongjun-ANU/MAUVE_MUSE_Zgas_SigmaSFR}.



\bibliographystyle{mnras}
\bibliography{reference} 




\appendix

\section{Global scaling-relation context of 14 MAUVE--MUSE galaxies}
\label{app:global_scaling}

\begin{figure*}
    \centering
    \includegraphics[width=\textwidth]{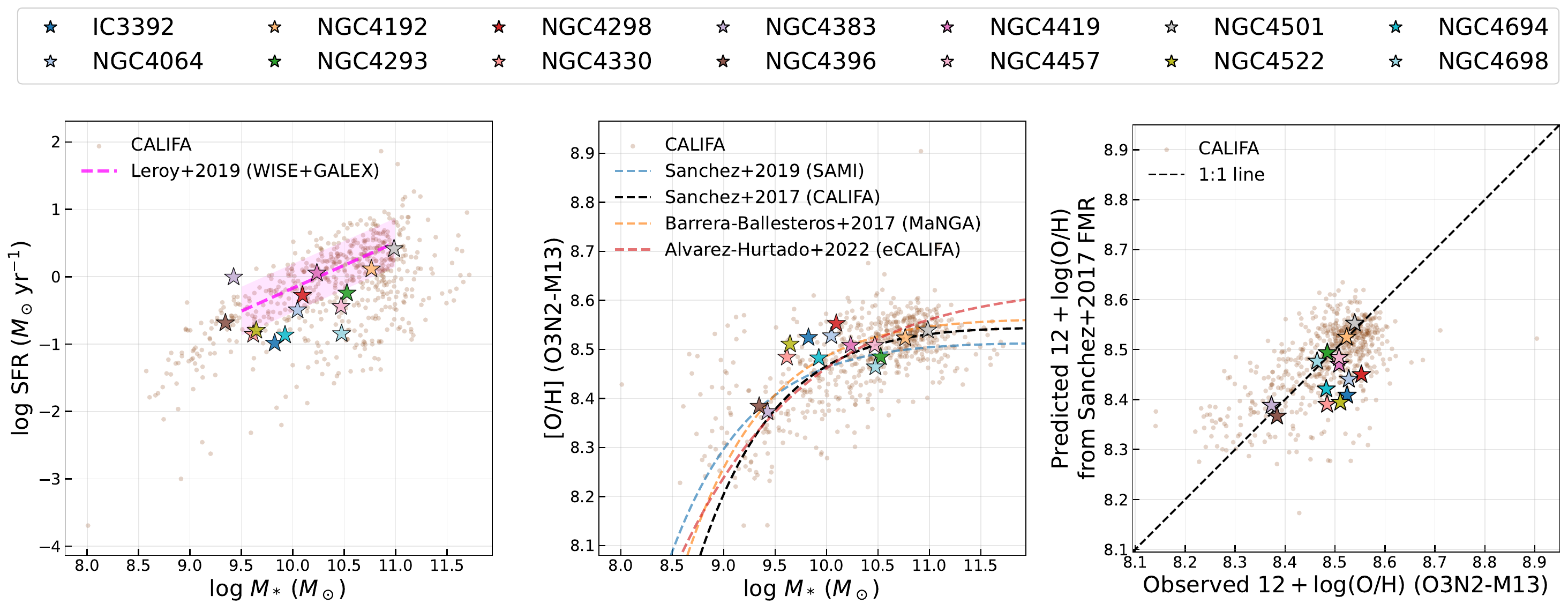}
    \caption{Global scaling-relation context of the 14 MAUVE--MUSE galaxies. \textit{Left:} integrated SFR--$M_*$ plane, compared with the local-universe SFMS (magenta dashed line with shaded region indicating the width of main sequence) from \citet{leroy19}. \textit{Middle:} integrated MZR, where MAUVE--MUSE galaxies are compared with several local reference relations, including \citet{sanchez17,sanchez19,barreraballesteros17,alvarezhurtado22}. \textit{Right:} observed integrated metallicity versus the metallicity predicted from the \citet{sanchez17} FMR plane, with the 1:1 relation indicated by the dashed line. In each panel, CALIFA galaxies are shown in the background.}
    \label{fig:sfms_mzr_fmr_global}
\end{figure*}

To place the MAUVE--MUSE galaxies in the context of global scaling relations, we compare their integrated properties with local-universe SFMS, MZR, and FMR trends in \autoref{fig:sfms_mzr_fmr_global}. Integrated stellar masses and star formation rates are adopted from \citet{leroy19}, rescaled to our assumed distance of 16.5 Mpc \citep{mei07}. Integrated gas-phase metallicities are measured by collapsing all valid MAUVE--MUSE spaxels in each galaxy, correcting the integrated line fluxes via the Balmer decrement as described in \autoref{sec:sfr}, and applying the \marino\ calibration. For comparison, we also show CALIFA galaxies in the background of all three panels, together with representative literature relations for the SFMS, MZR, and FMR.

The left panel shows that several MAUVE--MUSE galaxies lie below the local SFMS, consistent with environmentally suppressed star formation in the Virgo cluster. By contrast, NGC~4383 lies above the sequence, in line with its starburst/outflow nature \citep{watts24,cortese26}. The middle panel shows that the galaxies remain broadly consistent with local integrated MZR trends, although some objects lie slightly above the fiducial relations. The right panel compares the observed integrated metallicities with those predicted from the \citet{sanchez17} FMR. Overall, the MAUVE--MUSE sample remains close to the global FMR relation, with a scatter of 0.055 dex, smaller than the 0.067 dex reported for the CALIFA sample in \citet{sanchez17}. In particular, NGC~4383 is not an outlier in the global MZR or FMR planes, indicating that its unusual behaviour discussed in the main text is primarily a local, rather than global, effect.

\bsp	
\label{lastpage}
\end{document}